\documentclass[useAMS,usenatbib]{mn2e}
\usepackage{epsfig}
\usepackage{color}
\usepackage{subfigure}
\usepackage{amsmath}
\usepackage{amssymb}

\def\lsim{\mathrel{\rlap{\lower3.5pt\hbox{\hskip0.5pt$\sim$}}
    \raise0.5pt\hbox{$<$}}}
\def\gsim{~\rlap{$>$}{\lower 1.0ex\hbox{$\sim$}}}

\newcommand{\goodgap}{\hspace{\subfigtopskip} \hspace{\subfigbottomskip}}

\title[DM and stellar accelerations]{Newtonian acceleration scales in spiral galaxies}
\author[V.F. Cardone \& A. Del Popolo]{V.F. Cardone$^{1}$\thanks{Corresponding author\,: {\tt winnyenodrac@gmail.com}}, A. Del Popolo$^{2,3}$ \\
$^1$I.N.A.F. - Osservatorio Astronomico di Roma, via Frascati 33, 00040 - Monte Porzio Catone (Roma), Italy \\
$^2$Dipartimento di Fisica e Astronomia, Universit\`{a} di Catania, Viale Andrea Doria 6, 95125 Catania, Italy \\
$^3$Departamento de Astronomia, Universidade de S\~ao Paulo,  05508-900, S\~ao Paulo, SP, Brazil}

\date{Accepted xxx, Received yyy, in original form zzz}

\begin{document}

\maketitle

\begin{abstract}

We revisit the issue of the constancy of the dark matter (DM) and baryonic Newtonian acceleration scales within the DM scale radius by considering a large sample of late\,-\,type galaxies. We rely on a Markov Chain Monte Carlo (MCMC) method to estimate the parameters of the halo model and the stellar mass\,-\,to\,-\,light ratio and then propagate the uncertainties from the rotation curve data to the estimate of the acceleration scales. This procedure allows us to compile a catalog of 58 objects with estimated values of the $B$ band absolute magnitude $M_B$, the virial mass $M_{vir}$, the DM and baryonic Newtonian accelerations (denoted as $g_{DM}(r_0)$ and $g_{bar}(r_0)$, respectively) within the scale radius $r_0$ which we use to investigate whether it is possible to define a universal acceleration scale. We find a weak but statistically meaningful correlation with $M_{vir}$ thus making us argue against the universality of the acceleration scales. However, the results somewhat depend on the sample adopted so that a careful analysis of selection effects should be carried out before any definitive conclusion can be drawn.

\end{abstract}


\section{Introduction}

Galactic systems cover a wide range in size, mass, luminosity, colours and morphology so that it is almost obvious that finding a single model able to fully reproduce this rich phenomenology is doubtful. It is nevertheless somewhat surprising that some order in this wide parameter space may be created thanks to the existence of statistically meaningful correlations among galaxy structural and dynamical parameters. Textbook examples are represented by the Faber\,-\,Jackson (1976) and Tully\,-\,Fisher (1977) relations between velocity indicator and luminosity for early and late\,-\,type galaxies, respectively, the Fundamental Plane \citep{7S,DD87} between velocity dispersion, size and effective intensity and the photometric plane \citep{K00,G02} between photometric parameters of elliptical systems. Besides their usefulness as possible distance indicators, the existence of scaling relations among galaxy structural parameters is evidence that some hitherto not fully explained mechanism has taken place during the formation and evolutionary processes leading to the currently observed correlations. Finding and interpreting scaling relations has therefore recently become an active research field in the literature of galaxy properties.

Motivated by these considerations, \cite{KF04} described the dark halo density profile $\rho_{DM}$ of Sc\,-\,Im and dwarf spheroidal galaxies with the pseudo\,-\,isothermal profile, $\rho_{DM}(r) = \rho_0 r_0^2/(r^2 + r_0^2)$, to fit the rotation curve data of 55 systems. Under the maximal disk hypothesis, they found that the quantity $\mu_D = \rho_0 r_0$, proportional to the halo central density, stays remarkably constant at the value $\mu_D \simeq 100 \ {\rm M_{\odot}/pc^2}$ for all the galaxies in their sample. Donato et al. (2009, hereafter D09) then further extended this result considering the coadded rotation curves of $\sim 1000$ spiral galaxies, individual mass models for spiral and dwarf systems and the weak lensing data of a set of elliptical galaxies. Adopting a \cite{B95} model for the dark halo, they found $\rho_0 r_0 = 141_{-52}^{+82} \ {\rm M_{\odot}/pc^2}$ over 14 units of $B$ magnitude (i.e., $14/2.5 = 5.6$ orders of magnitude in luminosity) where it is important to stress that $(\rho_0, r_0)$ are defined with respect to the Burkert rather than pseudo\,-\,isothermal profile. The D09 result may also be expressed in terms of the Newtonian acceleration at the scale radius $r_0$. This is given by\,:

\begin{displaymath}
g_{DM}(r_0) = \frac{G M_{DM}(r_0)}{r_0^2} = 3.2^{+1.8}_{-1.2} \ \times \ 10^{-9} \ {\rm cm/s^2}
\end{displaymath}
and should stand out as a universal quantity provided that D09 result is correct. This result has since been generalized to the gravitational acceleration of the baryons within $r_0$ by Gentile et al. (2009, hereafter G09). Using a sample of spiral galaxies with carefully measured rotation curves and excluding some systems based on quality criteria, G09 found

\begin{displaymath}
g_{bar}(r_0) = 5.7^{+3.8}_{-2.8} \ \times \ 10^{-10} \ {\rm cm/s^2}
\end{displaymath}
to be constant for all the galaxies in their sample.

As already stated above, both the D09 and G09 results were obtained adopting the Burkert density profile,

\begin{displaymath}
\rho(r) = \frac{\rho_0 r_0^3}{(r+r_0)(r^2+r_0^2)}
\end{displaymath}
for the DM haloes. According to D09, this choice is motivated by the conducive property of the Burkert profile to mimic a pseudo\,-\,isothermal core in the inner regions and a $r^{-3}$ scaling in the outer ones, similar to the popular NFW profile \citep{NFW}. Moreover, such a model is known to make extremely good fits to rotation curve data \citep{G04,G07,DeP09} although some few remarkable exceptions exist \citep{Simon,DeP12}.

The D09 and G09 results have also been revised and criticized by different authors. First, \cite{B09} extended the analysis to group and cluster scale systems evaluating the column density defined as\,:

\begin{displaymath}
S = \frac{2}{r_{\star}^2} \int^{r_\star}_{0}{r dr \int{dz \rho_{DM}(\sqrt{r^2+z^2})}}
\end{displaymath}
which is proportional to the mean surface density within $r_{\star}$ (given by $\rho_{\star} r_{\star}$) and exactly equal to $\rho_0 r_0$ for the Burkert model. A fit to their sample data gives

\begin{displaymath}
\log{S} = 0.21 \log{\left ( \frac{M_{halo}}{10^{10} \ {\rm M_{\odot}}} \right )} + 1.79
\end{displaymath}
with $S$ in ${\rm M_{\odot}/pc^2}$ thus contradicting the D09 results on the constancy of $\rho_0 r_0$. Similarly, \cite{NRT10} considered a large sample of elliptical galaxies and estimated the quantity $\langle \rho_{DM} \rangle R_{eff}$ which is proportional to the column density defined by \cite{B09}, but using the effective radius $R_{eff}$ as reference scale. Although they agree with D09 that this quantity is constant for spirals, these authors nevertheless found that early\,-\,type galaxies violate the constant density scenario by a factor $\sim 10$ on average and by a factor $\sim 5$ if one only considers systems spanning the same mass regime. A further analysis has been carried out by \cite{CT10} based on strong lensing and velocity dispersion data for intermediate redshift lens galaxies. They found that the column density correlates with the mass and the luminosity although the slope of the scaling relations depend on both the DM halo model adopted and the stellar IMF.

The original D09 and G09 results and the following tests quoted above can not be straightforwardly compared because of the differences in the reference radii adopted, the halo model and the mass and luminosity ranges covered. As such, drawing a definitive conclusion on the existence or not of a universal quantity is actually not an easy task. In an attempt to give a clear cut answer to this question, we revisit here the D09 and G09 results by mimicking as close as possible their original analysis, but improving some substantial points. First, we almost double the G09 sample which also allows us to investigate possible selection effects. Second, we estimate the Burkert halo model parameters by directly fitting the rotation curve data rather than adjusting previous results already available in the literature and based on the use of different density profiles. Third, we adopt a Bayesian fitting procedure to infer a realistic estimate of the errors on the quantities of interest also taking into account the uncertainties on the mass\,-\,to\,-\,light ratio. Rather than considering a column density, we express our results in terms of the Newtonian acceleration at the DM scale radius and investigate whether this quantity correlates with the luminosity, mass and size of the galaxies.

The plan of the paper is as follows. Sect.\,2 gives some general consideration on how we define and compute the Newtonian acceleration also explaining to what extent this quantity may or may not be considered model independent. We then present the rotation curve sample in Sect.\,3 and discuss the results of fitting our adopted galaxy model (including gas, disc and DM  components) to the data. Scaling relations are investigated in Sect.\,4, while our summary and discussion are finally presented in Sect.\,5. In Appendix A, we list the constraints on the model parameters and the estimate of the Newtonian accelerations so that the interested reader can use them for any further investigation.

\section{Newtonian acceleration scales}

The concept of Newtonian acceleration is so well known that spending more than few words on it could seem quite superfluous. Nevertheless, while its definition is rather simple, some subtleties arise when dealing with actual galaxies. First, spiral galaxies may be considered three component systems with stars and gas representing the baryon mass and the halo accounting for the dark matter term. In a first and reasonably good approximation, stars are distributed in a thin disc well modeled by the Freeman (1970) exponential law, while the gas component roughly follows the disc but with strong irregularities eventually leading to a clumpy distribution. Modeling such an irregular profile with an analytical expression could introduce a bias in the derivation of the Newtonian acceleration thus altering the analysis of the scaling relations we are interested in. We therefore use a different approach noting that, for a spherically symmetric system, the circular velocity simply reads

\begin{displaymath}
v_c^2(r) = G M(r)/r \ .
\end{displaymath}
Since observations of the luminosity profile directly provide us the circular velocity of both baryonic components, we can define an effective mass as

\begin{displaymath}
M_{eff}(r) = r v_c^2(r)/G
\end{displaymath}
and then estimate the corresponding acceleration as

\begin{figure*}
\centering
\subfigure{\includegraphics[width=5.25cm]{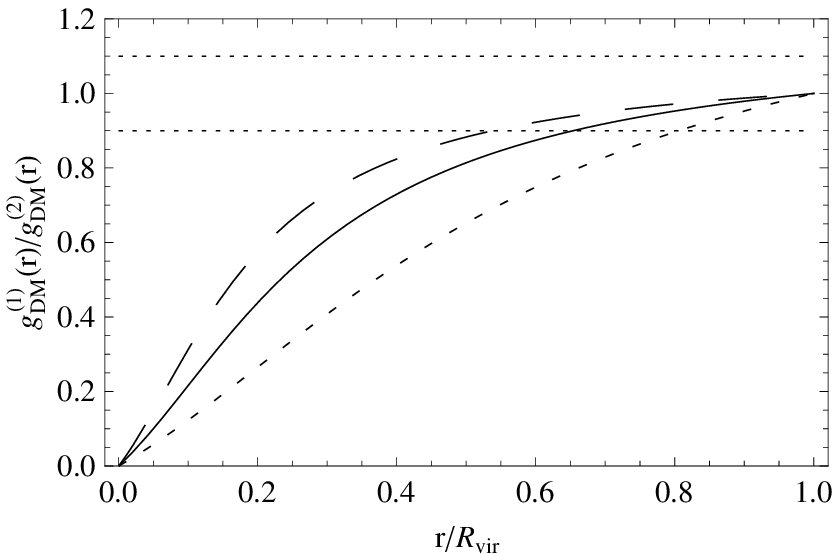}} \goodgap
\subfigure{\includegraphics[width=5.25cm]{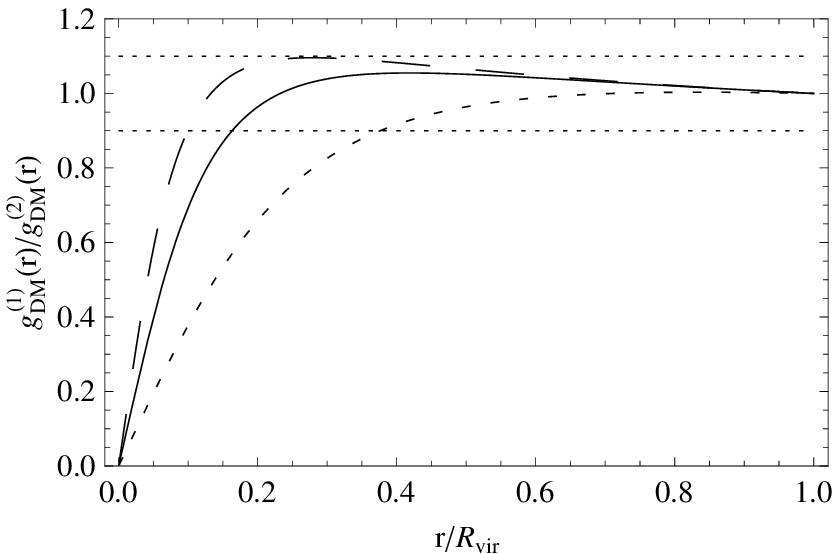}} \goodgap
\subfigure{\includegraphics[width=5.25cm]{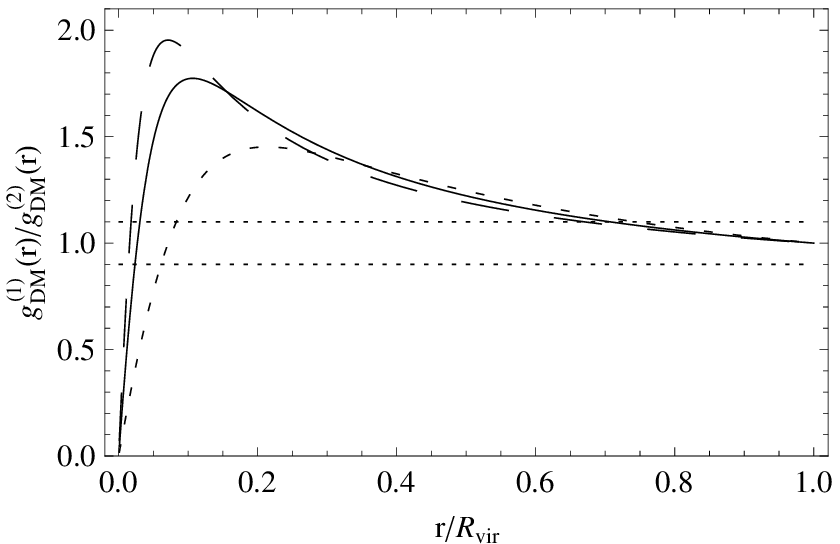}} \goodgap
\caption{Dark matter acceleration ratio for the Burkert vs NFW models assuming the same virial velocity. Short dashed, solid and long dashed lines refer to models with $c_{NFW} = 2.5, 5.0, 7.5$, respectively, while we set $c_{2}(Bur)/c_2(NFW) = 0.5, 1.0, 2.0$ from left to right panels. Dotted lines delineate the region where the ratio is equal to unity within a $10\%$ confidence.}
\label{fig: dmaccratio}
\end{figure*}

\begin{displaymath}
g(r) = G M_{eff}(r)/r^2 \ .
\end{displaymath}
We therefore define the Newtonian accelerations as\,:

\begin{equation}
\left \{
\begin{array}{l}
\displaystyle{g_{disc}(r) = \frac{G M_{disc}(r)}{r^2} = \frac{v_{disc}^2(r)}{r}} \\
~ \\
\displaystyle{g_{gas}(r) = \frac{G M_{gas}(r)}{r^2} = \frac{v_{gas}^2(r)}{r}} \\
\end{array}
\right .
\label{eq: discgasacc}
\end{equation}
where $v_{disc}$ and $v_{gas}$ are the measured stellar disc and gas circular velocities at the radial distance $r$ from the galaxy centre. It is worth stressing that Eqs.(\ref{eq: discgasacc}) are exact only for a spherical mass distribution, while neither the disc nor the gas are spherical systems. Although not being formally correct, such definitions possess the remarkable property to be directly related to the data thus avoiding the errors due to an incorrect modeling of the mass distribution related to gradients in the stellar $M/L$ ratio or clumpiness in the gas profile.We nevertheless stress that, when fitting rotation curve data, we do not rely on the approximated Eqs.(\ref{eq: discgasacc}), but rather correctly compute the theoretical circular velocity from the gravitational potential obtained by solving the Poisson equation for the given component\footnote{Actually, for most cases, the original references for the data already provide also the circular velocity of the disc and gas components assuming a fiducial stellar $M/L$ ratio $\Upsilon$ so that we have only to rescale them to match our estimated $\Upsilon$.}. Should we have used the spherical approximation, significant biases could have been introduced in the estimate of the halo parameters depending on the details of the disc and gas distribution properties (such as, e.g., the disc scalelength and the gas clumpiness).

Computing the Newtonian acceleration for the dark matter term presents a different problem. On the one hand, it is customary to assume that the halo is spherically symmetric so that the acceleration is simply $g_{DM}(r) = G M_{DM}(r)/r^2$ with $M_{DM}(r)$ the halo mass within $r$. On the other hand, the halo circular velocity is not directly measured so that one can not resort to the analog of Eq.(\ref{eq: discgasacc}) to infer the DM acceleration from the data. One is therefore forced to assume a halo model, determine its parameters by fitting to the observed total rotation curve and then estimate the acceleration. In order to be consistent with G09, we will assume a \cite{B95} model so that the mass profile reads\,:

\begin{equation}
M_{DM}(r) = M_{vir} \ \times \ \frac{\mu(r)}{\mu(R_{vir})}
\label{eq: burmass}
\end{equation}
with $M_{vir}$ and $R_{vir} = [(3 M_{vir})/(4 \pi \Delta_{vir} \rho_{crit})]^{1/3}$ the virial mass and radius\footnote{We follow \cite{BN98} to set $\Delta_{vir}$, while $\rho_{crit} = 3 H_0^2/8 \pi G$ is the critical density of the universe.}, respectively, and

\begin{equation}
\mu(r) = \ln{\left ( 1 + \frac{r}{r_0} \right )} - \arctan{\left ( \frac{r}{r_0} \right )}
+ \frac{1}{2} \ln{\left [ 1 + \left ( \frac{r}{r_0} \right )^2 \right ]}
\label{eq: mubur}
\end{equation}
with $r_0$ a characteristic radius. The model is assigned by the two parameters $(M_{vir}, r_0)$, but, when fitting to the rotation curve data, it is more convenient to reparameterize the model in terms of the circular velocity at the virial radius, $V_{vir}^2 = G M_{vir}/R_{vir}$, and the concentration $c_{vir} = R_{vir}/r_0$.

The DM Newtonian acceleration can then be written as

\begin{equation}
g_{DM}(r) = \frac{G M_{vir}}{r^2} \frac{\mu(r)}{\mu(R_{vir})} = g_{vir} \ \times \ \frac{\mu(r)/r^2}{\mu(R_{vir})/R_{vir}^2}
\label{eq: dmacc}
\end{equation}
having set $g_{vir} = G M_{vir}/R_{vir}^2 = V_{vir}^2/R_{vir}$ for the acceleration at the virial radius. Eq.(\ref{eq: dmacc}) allows us to immediately estimate how the acceleration depends on the halo model. First, we note that, since the virial velocity $V_{vir}$ is typically well determined by fitting the rotation curve, its value can be taken as model independent. That is to say, in order to have two different models fitting the same data, they must have the same virial velocity. We will therefore consider two different models with the same $V_{vir}$ (and hence the same $R_{vir}$ and $g_{vir}$ values), but two different mass profiles described by the functions $\mu_1(r)$ and $\mu_2(r)$. The ratio among the corresponding accelerations will then be\,:

\begin{equation}
\frac{g_{DM}^{(1)}(r)}{g_{DM}^{(2)}(r)} = \frac{\mu_1(r)/\mu_1(R_{vir})}{\mu_2(r)/\mu_2(R_{vir})} \ .
\label{eq: dmaccratio}
\end{equation}
As an example, we consider here the Burkert and NFW \citep{NFW} as models 1 and 2, respectively, thus setting

\begin{equation}
\mu_2(r) = \ln{\left ( 1 + \frac{c_{NFW} r}{R_{vir}} \right )} - \frac{c_{NFW} r}{R_{vir}} \left ( 1 + \frac{c_{NFW} r}{R_{vir}} \right )^{-1}
\label{eq: munfw}
\end{equation}
with $c_{NFW} = R_{vir}/r_s$ and $r_s$ the characteristic radius of the NFW profile. In order to meaningfully compare the models, we first introduce a new concentration parameter defined as $c_{2} = R_{vir}/R_{2}$ with $R_{2}$ the radius where the logarithmic slope of the density profile, $\alpha = -d\ln{\rho}/d\ln{r}$, equals the isothermal value $\alpha = -2$. It is easy to check that

\begin{displaymath}
c_{2} = \frac{R_{vir}}{R_2} = \left \{
\begin{array}{ll}
\displaystyle{\frac{R_{vir}}{1.52 r_0} = \frac{c_{vir}}{1.52}} & {\rm for \ Burkert} \\
~ & ~ \\
\displaystyle{\frac{R_{vir}}{R_s} = c_{NFW}} & {\rm for \ NFW} \\
\end{array}
\right . \ .
\end{displaymath}
Fig.\,\ref{fig: dmaccratio} shows the ratio $g_{DM}^{(1)}(r)/g_{DM}^{(2)}(r)$ for the Burkert vs NFW models as a function of $r/R_{vir}$ assuming the two models have the same virial velocity and different values of the NFW concentration $c_{NFW}$ and $c_{2}(Bur)/c_{2}(NFW)$. This plot highlights an important feature of the Newtonian acceleration, namely its dependence on the particular halo model adopted. This point clearly has to be taken into account when looking for scaling relations among $g_{DM}(r)$ and stellar or DM quantities. The central panel, however, suggests a possible way out of this problem. Indeed, if we ask that two different models fit the same rotation curve data, their circular velocity profiles must be similar, at least over the radial range probed by the data. We have checked by trial and error that, for the Burkert and NFW models, this is the same as asking that the virial velocity $V_{vir}$ is the same and the $c_2$ concentration is comparable. In such a case, the central panel in Fig.\,\ref{fig: dmaccratio} shows that $g_{DM}^{Bur}(r) \simeq g_{DM}^{NFW}(r)$ within $10\%$ if $r > r_{min}$ with $r_{min} \sim 0.1 - 0.2 R_{vir}$ (depending on the $c_{NFW}$  value). In the following, we will use the Burkert model to estimate the acceleration values, but will investigate scaling relations only for values of $r > r_{min}$ in order to be confident that a mismatch in the modeling does not significantly bias the results.

\section{Estimating the acceleration}

Both the baryon and DM accelerations require rotation curve data in order to be evaluated. Indeed, HI observations directly give the amount of gas and its profile thus allowing us to infer $v_{gas}(r)$, while photometry and an estimate of the stellar $M/L$ ratio gives $v_{disc}(r)$. Adding in quadrature these terms to the theoretically predicted DM circular velocity allows us to get the combined circular velocity to be compared with the rotation curve and hence determine the halo model parameters $(V_{vir}, c_{vir})$. In the following, we will first describe the galaxy sample we used and the fit procedure and then will explain how we estimated the acceleration scales.

\subsection{Data and fitting method}

We have searched the literature for systems with high quality rotation curve data probing the gravitational potential with good sampling, large radial extent (i.e., up to $R > R_{opt} = 3.2 R_d$ with $R_d$ the disk scale length) and small errors. Whenever possible, we rely on H$\alpha$ data or a combination of HI and H$\alpha$ so that the potential impact of beam smearing on the circular velocity is reduced and does not bias the estimate of the halo parameters\footnote{A possible concern on the use of the H$\alpha$ data is related to H$\alpha$ possibly tracing biased star forming regions of the disc (as, e.g., spiral arms). As a consequence, the measured circular velocity could be non representative of the disc stars kinematics. While we can not definitively exclude such a possibility, it is nevertheless worth noting that, whenever both HI and H$\alpha$ data are available in the same region, they closely track each other. We are therefore confident that no significant bias is induced by the use of H$\alpha$ data}. We finally end up with 58 galaxies from different sources \citep{dBB02,Simon,Spano,Things,Oh,Sw11} spanning nine units of $B$ band magnitude (roughly, $-22 \le {\cal{M}}_B \le -13$) and 9/2.5 orders in luminosity and different spiral classes (from dwarfs system to low and high surface brightness galaxies).

It is worth stressing that our compilation is not based on any selection criteria other than the availability of measurements covering a large radial range. One could, however, wonder whether such a blind selection may lead to reliable results. Indeed, poor quality curves may be fitted by models with unreliable parameters so that the estimate of the acceleration scales should not be trusted. On the other hand, selecting a priori only a subset of the galaxies may risk bias in the analysis of scaling relations excluding systems which drive an eventual weak correlation. In order to investigate this issue, we will start by retaining all the galaxies and then look for the impact of different quality cuts on the slope and scatter of the scaling relations among acceleration scales and other parameters of interest.

In order to determine the model parameters, we carry out a likelihood analysis by maximizing ${\cal{L}}({\bf p}) \propto \exp{[-\chi^2({\bf p})/2]}$ with

\begin{equation}
\chi^2 = \sum_{i = 1}^{{\cal{N}}}{\left [ \frac{v_c^{obs}(r_i) - v_{c}^{th}(r_i, {\bf p})}{\sigma_i} \right ]^2}
\label{eq: defchisq}
\end{equation}
where $v_c^{obs}(r_i)$ is the measured circular velocity for the $i$\,-\,th point (with $\sigma_i$ the error), $v_c^{th}(r_i, {\bf p})$ the theoretically predicted value for the given set of parameters ${\bf p}$ and the sum is over the ${\cal{N}}$ data points. It is worth stressing that, since the errors reported from different authors have been obtained in different ways and typically take care of possible systematic effects, the reduced $\chi^2$ for the best fit model is not expected to be unity.

The fitting procedure allows us to determine the halo model parameters $(V_{vir}, c_{vir})$ and the stellar $M/L$ ratio $\Upsilon_{\star}$. This could seem at odds with our previous statement that the disc circular velocity is directly estimated from the data. Actually, this is only approximately true. Indeed, the photometry of the galaxies allows to set the disc profile and hence evaluate the corresponding circular velocity provided $\Upsilon_{\star}$ is known. This latter quantity can still be guessed from the multiband photometric data using a prescription to convert colors into a fiducial $\Upsilon_{\star}$. However, uncertainties in the properties of the stellar population (such as, e.g., its age and metallicity) makes the estimated $\Upsilon_{\star}$ affected by a large error. To take care of this issue, rotation curve papers typically report the values of $v_{disc}(r)$ for $\Upsilon_{\star} = 1$ so that the actual stellar $M/L$ ratio has to be determined by the fit to the data. We therefore end up with a three parameters set, i.e., in Eq.(\ref{eq: defchisq}), it is ${\bf p} = (\Upsilon_{\star}, V_{vir}, c_{vir})$.

\begin{figure*}
\centering
\subfigure{\includegraphics[width=5.25cm]{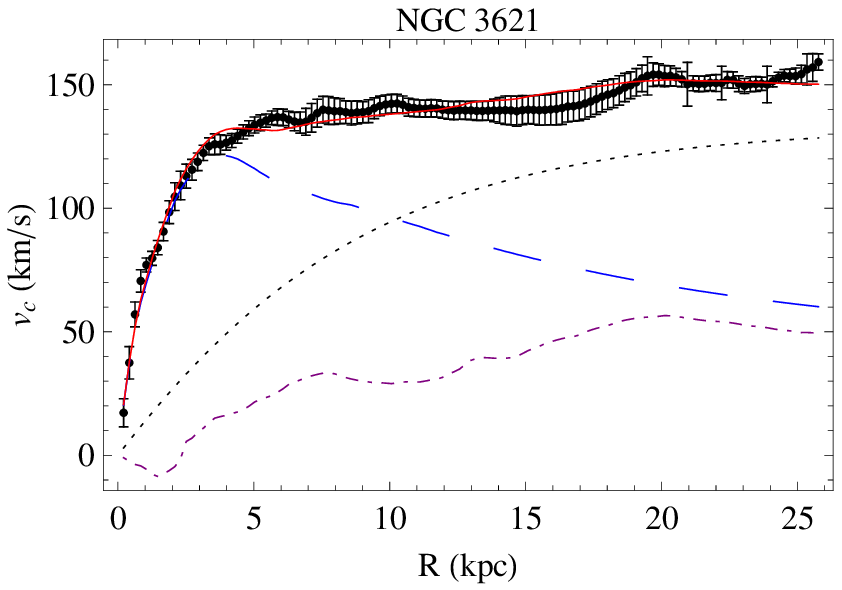}} \goodgap
\subfigure{\includegraphics[width=5.25cm]{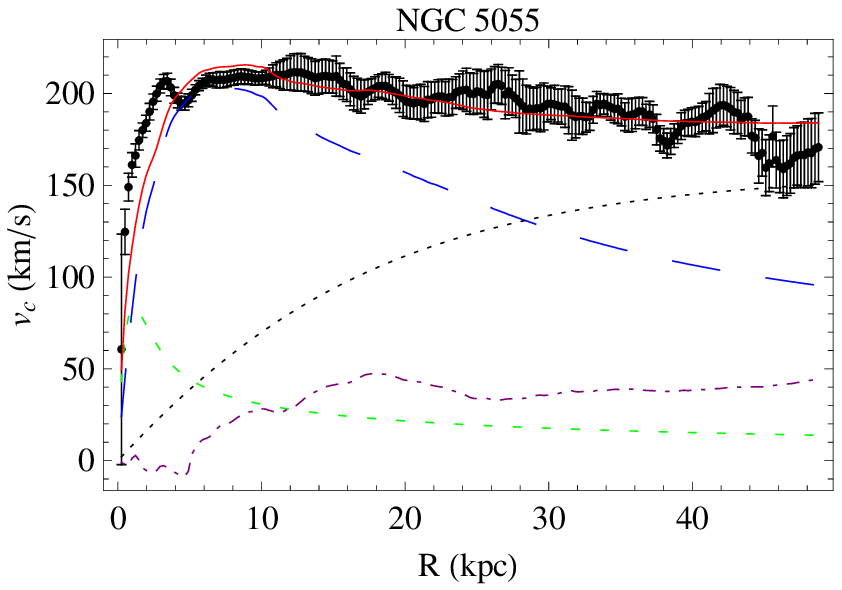}} \goodgap
\subfigure{\includegraphics[width=5.25cm]{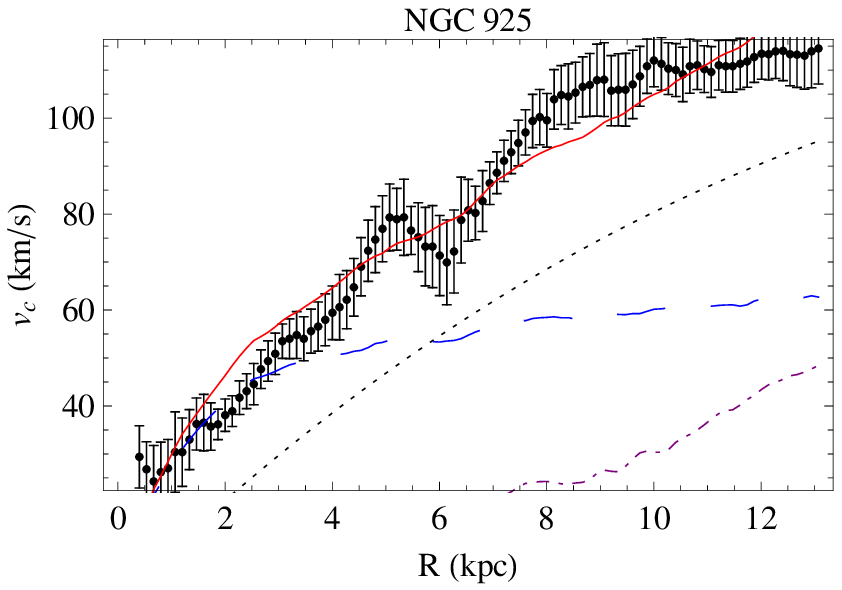}} \goodgap
\caption{Best fit models superimposed on the data for three example galaxies. Red solid, blue long dashed, purple dot dashed, black dotted and green short dashed lines refer to the the full, disc, gas, halo and bulge contributions to the the circular velocity.}
\label{fig: rcex}
\end{figure*}

In order to efficiently explore this 3D space, we use a Markov Chain Monte Carlo (MCMC) method running three chains checking the convergence through the \cite{GR92} criterion. After cutting the chains to avoid the burn\,-\,in phase and thinning them to erase spurious correlations, we merge the three chains in a single one which can be used to infer constraints on the model parameters. Note that, in a Bayesian framework, the best fit parameters ${\bf p}_{bf}$ do not represent the most reliable estimate of the single parameters $p_i$. On the contrary, one should look at the marginalized likelihoods obtained by integrating ${\cal{L}}({\bf p})$ over all the parameters but the $i$\,-\,th one. Using the MCMC method, this reduces to look at the histogram of the $i$\,-\,th parameter values to infer the median and the $68$ and $95\%$ confidence ranges. Although we will report also the best fit, in the following, we will consider the median and $68\%$ confidence range as our final estimates of the model parameters.

\subsection{Results from rotation curve fitting}

It has long been known that cored models are quite efficient at fitting spiral galaxies rotation curves. It is therefore not surprising that the adopted Burkert profile provides a good fit for almost all of the galaxies in our sample. It is nevertheless worth spending some more words on how one can decide whether the match between the data and the best fit model can be considered as good or not. As already stressed above, the errors on the measured  circular velocity are not Gaussian distributed since they can include contributions for the asymmetry between the approaching and receding sides of the galaxy. As a first consequence, one can not rely on the usual reduced $\chi^2$ criterion (i.e., $\tilde{\chi}^2 = \chi^2/d.o.f. \simeq 1$) to deem a fit as good or bad. Moreover, our sample is comprised of galaxies collected from different authors, each one of them using their own recipe for assigning errors to the rotation curve data. As such, it is not straightforward to work out a single quantitative criterion which works equally well for all cases. We have therefore relied on three different qualitative criteria to deem the matching among data and best fit model as evidence in favour of the Burkert profile. First, we look at the $\tilde{\chi}^2$ value. Indeed, although its normalization depends on how the errors have been evaluated, it is nevertheless clear that a large value is indicative of a poor fit (unless the errors have been grossly underestimated which is not the case). A small $\tilde{\chi}^2$ value can nevertheless be obtained if the errors have been overestimated. As a consistency check, we will therefore look at the root mean square of the residuals and their trend with the radius. For a model well fitting the data, we indeed expect that the residuals (scaled with respect to the measurement errors or to the observed circular velocity values) are small and scatter around the mean value. As such, the rms of the residuals will be small and no correlation with $r$ can be observed. On the contrary, a trend with $r$ will be obtained if the model is systematically failing to fit the data.

A few examples will help to clarify the use of the above quality criteria. The left panel in Fig.\,\ref{fig: rcex} shows the rotation curve data for NGC\,3621 and the superimposed best fit model ($\tilde{\chi}^2 = 0.75$). This is a typical case where one could rely on the expectation based on the reduced $\chi^2$ value are in accordance with our definition of {\it well fitted system}. The best fit model closely tracks the data and there is no trend of the very small residuals with the distance from the galaxy centre. On the contrary, the case of NGC\,5055 (shown in the central panel of Fig.\,\ref{fig: rcex}) is an example of a system with a large reduced $\chi^2$ ($\tilde{\chi}^2 = 3.63$), but with a nevertheless reasonably well fitted rotation curve. Indeed, the best fit model tracks the data almost everywhere but in the very inner regions where the circular velocity is likely to be mainly determined by the bulge component which we hold fixed in our analysis. We have therefore included this galaxy in our sample of {\it well fitted systems} since the divergences may be very likely an evidence of a failure in the bulge rather than the halo modeling. Finally, the right panel of Fig.\,\ref{fig: rcex} shows the data and the best fit model for NGC\,925 which is a good example of why one must not only rely on $\tilde{\chi}^2$ value. Indeed, we get $\tilde{\chi}^2 = 1.31$ for this galaxy (smaller than for NGC\,5055), but the best fit model fails to follow the data overestimating the circular velocity in the inner regions and underestimating it in the outer ones. Moreover, the best fit circular velocity profile keeps almost linearly increasing, whereas the data shows some hints of a flattening of the rotation curve. Motivated by these discrepancies, we therefore consider this fit an example case of a {\it poorly fitted system}.

Applying these conservative criteria, we finally find that all the galaxies but six are well fitted by the Burkert profile. This is in agreement with the previous results in the literature which have shown that this halo model does a very good job in fitting the rotation curves of spiral galaxies of different morphological classes. Although this could be considered as further evidence in favour of cored vs cuspy models, it is worth stressing that tackling this unresolved debate is not our aim here. From our point of view, having successfully fitted the rotation curve data of such a large sample of spiral galaxies allows us to confidently rely on the Burkert profile to estimate the Newtonian acceleration scales of interest here.

\subsection{Acceleration scales}

Having constrained the disc stellar $M/L$ ratio and the halo model parameters, we can now use Eqs.(\ref{eq: discgasacc}) and (\ref{eq: dmacc}) to estimate the Newtonian acceleration of both the baryon (disc\,+\,gas) and DM components. To fully take into account the uncertainties on the fitting parameters and the correlations among them, we evaluate $g_i(r_j)$ along the Markov chain for each galaxy and use the resulting values to estimate the median and $68\%$ confidence ranges. Note that, here, the label $i$ denotes the galaxy component (baryons or DM), while $j$ refers to the reference radius adopted. Two choices are possible. First, one can set $r_j = R_d$, i.e. estimate the acceleration at the disc scalelength radius. On one hand, such a choice could be well motivated noting that the data mainly sample the intermediate disc regions (i.e., $R \sim few \ R_d$) so that $g_i(R_d)$ is likely to be well constrained. Actually, this is also the region where uncertainties on the baryon components (such as the disc stellar $M/L$ and the bulge modeling) have the larger impact. We therefore prefer to set $r = r_0$ in order to be consistent with the G09 analysis. It is worth stressing, however, that, for the typical concentration values we have obtained from the fitting analysis, one gets $r_0/R_{vir} < 0.1$ so that one is actually estimating the acceleration in the region which is most sensitive to changes in the halo model as can be seen from Fig.\,\ref{fig: dmaccratio}. We therefore warn the reader to not rely on any scaling rule to infer the $g_{DM}(r_0)$ values for a different model from the ones we find here using the Burkert profile.

\begin{figure*}
\centering
\subfigure{\includegraphics[width=7.5cm]{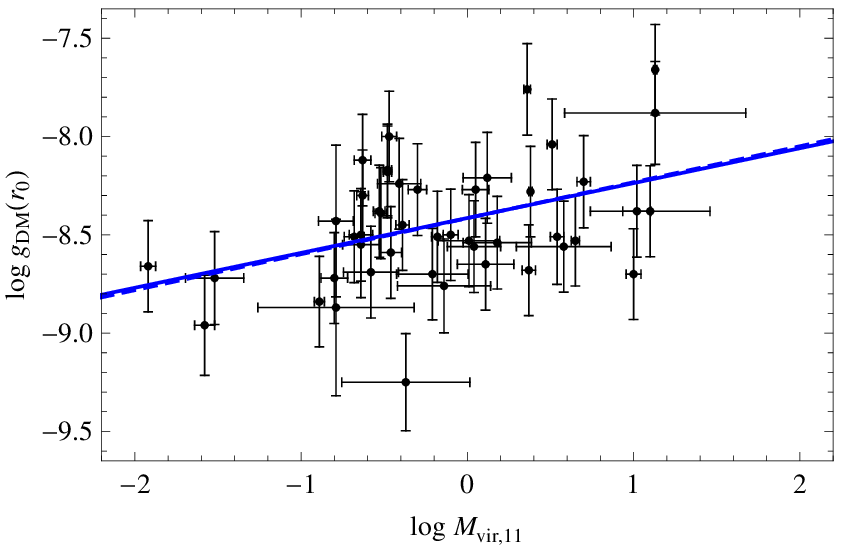}} \goodgap
\subfigure{\includegraphics[width=7.5cm]{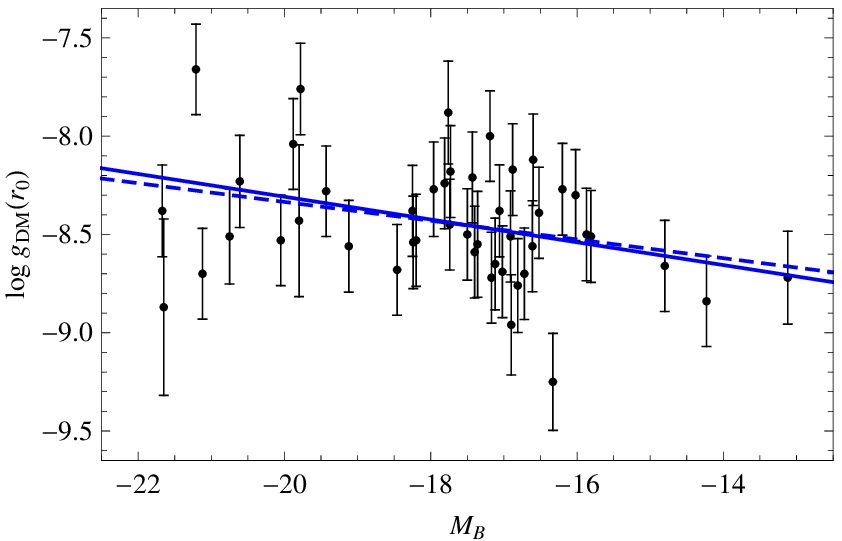}} \goodgap
\caption{Best fit models superimposed on the data for the DM acceleration $g_{DM}(r_0)$ as a function of the halo virial mass $\log{M_{vir,11}} = \log{(M_{vir}/10^{11} \ {\rm M_{\odot}})}$ and the galaxy absolute $B$ band magnitude $M_B$. The solid (dashed) line refers to the Bayesian (direct) fit.}
\label{fig: gdmfigs}
\end{figure*}

Although the data on the rotation curve typically extend up to large $r/R_d$ values, it is nevertheless possible that the constrained $r_0$ is larger than $R_{max}$, this latter quantity being the last measured point of the rotation curve. In such a cases, one could still rely on the estimated $r_0$ to infer the $g_i(r_0)$ values, but this comes at the price of assuming that the model fitted up to $R_{max}$ can be extrapolated outside. In order to avoid any possible systematic error related to such a choice, we prefer to exclude the galaxies with $r_0 > R_{max}$ even if their rotation curve are well fitted by the assumed model. For completeness, we give in Table,\,\ref{tab: gdmres} the constraints on $g_i(r_0)$ for all the galaxies in our sample, but we will use in the rest of the paper only those systems having both a well fitted rotation curve and $r_0 < R_{max}$ referred to in the following as the {\it Good} sample.

\section{Scaling relations}

The analysis first laid out in G09 (but see also \citealt{KF04,D09} for preliminary investigations) has concluded that both the baryon and the DM accelerations evaluated at the Burkert scale radius $r_0$ stay constant over almost 14 $B$ magnitudes in luminosity with $g_{bar}(r_0) = 5.7_{-2.8}^{+3.8} \times 10^{-10} \ {\rm cm/s^2}$ and $g_{DM}(r_0) = 3.2_{-1.2}^{+1.8} \times 10^{-9} \ {\rm cm/s^2}$ as inferred from a sample of 28 spiral galaxies along the Hubble sequence. Using our larger sample, we can revisit their results, but rely on a somewhat different procedure. First, we note that G09 did actually not perform any fit to the rotation curve data, but relied on the literature and an approximate law to convert the scale length of the model used by other authors to their Burkert model $r_0$ radius. While such a scaling law gives a correct local matching among different models, it is worth stressing that this does not automatically ensure that the theoretically predicted rotation curves well match also in the outer regions. On the contrary, here, we fit the full rotation curve with the Burkert model thus being also able to check whether or not the model well reproduces the observed data. As a result, we find that the scaling law used by G09 indeed provides values of $g_{i}(r_0)$ consistent with our ones with $\langle \Delta \log{g_{DM}(r_0)} \rangle = -0.03$ and  $\langle \Delta \log{g_{bar}(r_0)} \rangle = -0.02$ as estimated by the 18 galaxies common to both samples\footnote{We overlook 10 galaxies from the G09 sample because of different reasons. First, we do not use the six Milky Way dwarf satellites since the estimate of $g_i(r_0)$ has been obtained relying on a fit to the velocity dispersion data, while here we are fitting rotation curves which are not available for these systems. We also reject four more galaxies since we have been unable to retrieve the rotation curve data so that we can not estimate the stellar $M/L$ ratio and the halo model parameters.}. However, in order for such an agreement to hold, one has first to check that the fit to the data is good, which is not the case for the full sample. As such, some of the galaxies (see Table\,\ref{tab: gdmres}) included by G09 must actually be excluded in order not to bias the final results.

\begin{figure*}
\centering
\subfigure{\includegraphics[width=5.25cm]{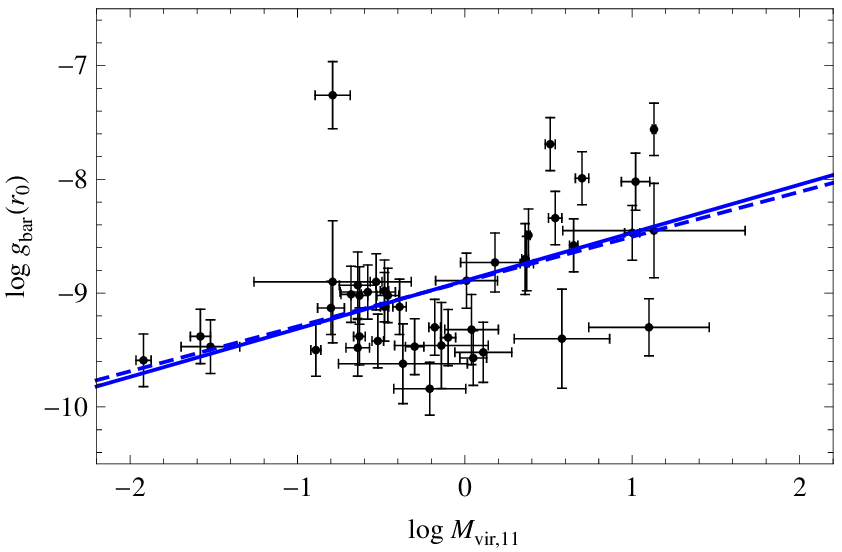}} \goodgap
\subfigure{\includegraphics[width=5.25cm]{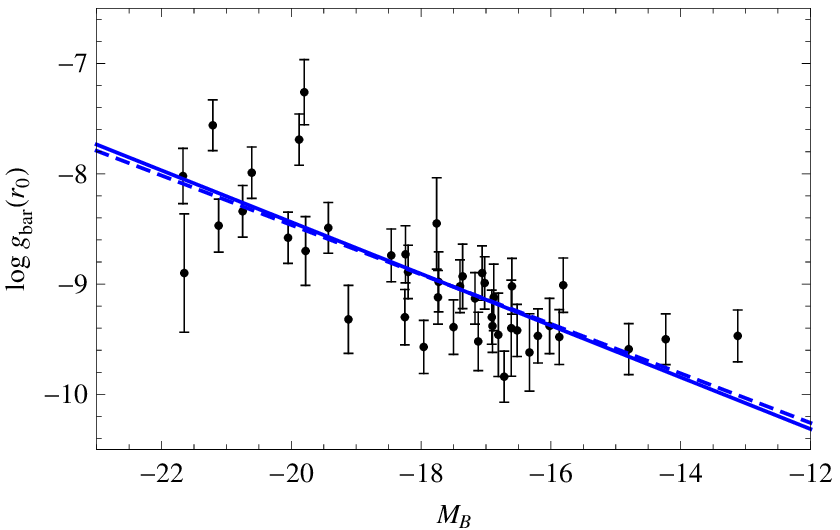}} \goodgap
\subfigure{\includegraphics[width=5.25cm]{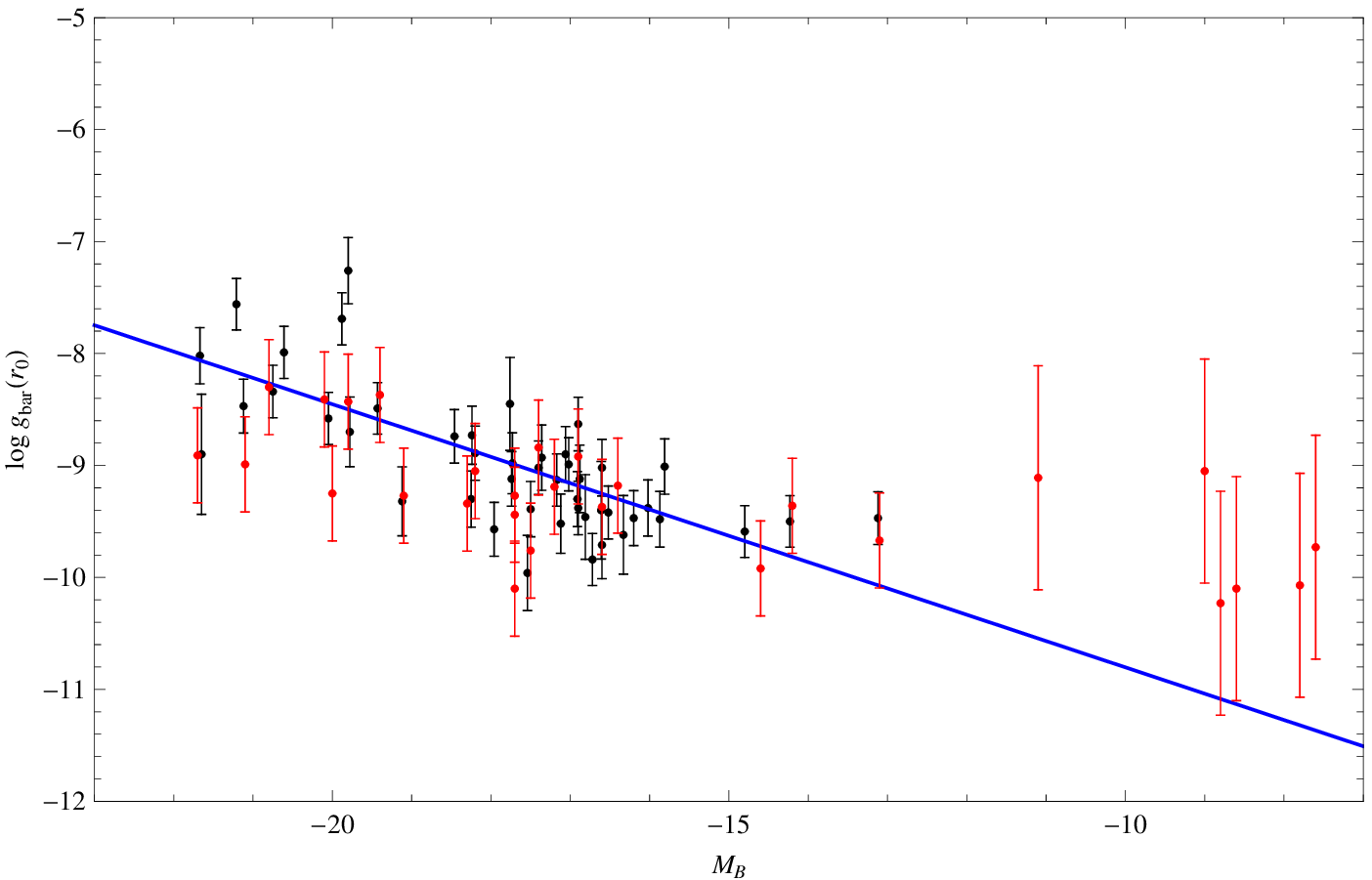}} \goodgap
\caption{{\it Left and centre.} Same as Fig.\,\ref{fig: gdmfigs} but for the baryon acceleration $g_{bar}(r_0)$. {\it Right.} Our best fit $g_{bar}(r_0)$ vs $M_B$ relation superimposed to the G09 (red) and our (black) acceleration data.}
\label{fig: gbarfigs}
\end{figure*}

\begin{table*}
\scriptsize
\caption{Scaling law parameters as output from the fit to three different galaxy subsamples. For each fit, we give the median and $68\%$ confidence range for the slope $a$, the intercept $b$ and the intrinsic scatter $\sigma_{int}$ obtained using the D'\,Agostini (2005) Bayesian method.}
\begin{center}
\begin{tabular}{cccccccccc}
\hline
~ & \multicolumn{3}{c}{Good} & \multicolumn{3}{c}{Common} & \multicolumn{3}{c}{Gold} \\
\hline
Id & $a$ & $b$ & $\sigma_{int}$ & $a$ & $b$ & $\sigma_{int}$  & $a$ & $b$ & $\sigma_{int}$ \\
\hline
$g_{DM}$\,-\,$M_{vir}$ & $0.166_{-0.051}^{+0.051}$ & $-8.416_{-0.007}^{+0.007}$ & $0.13_{-0.05}^{+0.05}$ & $0.059_{-0.030}^{+0.055}$ & $-8.428_{-0.003}^{+0.002}$ & $0.17_{-0.08}^{+0.10}$ & $0.131_{-0.047}^{+0.051}$ & $-8.414_{-0.008}^{+0.006}$ & $0.08_{-0.04}^{+0.06}$ \\
\multicolumn{10}{c}{~} \\
$g_{DM}$\,-\,$M_{B}$ & $-0.051_{-0.023}^{+0.017}$ & $-9.303_{-0.372}^{+0.212}$ & $0.14_{-0.06}^{+0.06}$ & $0.006_{-0.003}^{+0.005}$ & $-8.297_{-0.079}^{+0.076}$ & $0.15_{-0.07}^{+0.10}$ & $-0.049_{-0.017}^{+0.018}$ & $-9.346_{-0.200}^{+0.334}$ & $0.08_{-0.04}^{+0.06}$ \\
\hline
$g_{bar}$\,-\,$M_{vir}$ & $0.401_{-0.104}^{+0.102}$ & $-8.898_{-0.009}^{+0.008}$ & $0.44_{-0.06}^{+0.08}$ & $0.298_{-0.112}^{+0.114}$ & $-8.888_{-0.004}^{+0.004}$ & $0.16_{-0.09}^{+0.09}$ & $0.503_{-0.088}^{+0.090}$ & $-8.890_{-0.011}^{+0.025}$ & $0.34_{-0.06}^{+0.07}$ \\
\multicolumn{10}{c}{~} \\
$g_{bar}$\,-\,$M_{B}$ & $-0.232_{-0.028}^{+0.030}$ & $-12.96_{-0.49}^{+0.27}$ & $0.27_{-0.05}^{+0.07}$ & $-0.139_{-0.038}^{+0.072}$ & $-11.34_{-0.74}^{+0.43}$ & $0.09_{-0.05}^{+0.07}$ & $-0.231_{-0.027}^{+0.029}$ & $-13.04_{-0.038}^{+0.064}$ & $0.23_{-0.06}^{+0.06}$ \\
\hline
\hline
\end{tabular}
\end{center}
\label{tab: fitscale}
\end{table*}

A further difference with G09 concerns the estimate of the uncertainties. Since we have determined rather than assumed the stellar $M/L$ ratio and the halo model parameters, we have been able to attach to each $g_i(r_0)$ value an error fully taking into account both the fitting parameters uncertainties and their correlations. In order to be as conservative as possible, we then add in quadrature a systematic error related to the uncertainties on the method used to determine $g_{i}(r_0)$. Such a systematic is hard to ascertain so that we empirically rely on the comparison with the G09 values for the galaxies common to both samples and add in quadrature $0.23$ to the statistical errors on $\log{g_{i}(r_0)}$ since this is the rms of both components of $\Delta \log{g_{i}(r_0)}$. The final uncertainties turns out to be smaller than the ones adopted by G09 based on a qualitative analysis of the impact of the $M/L$ and systematic uncertainties. We are, however, confident that our estimate of the errors is better motivated by a statistical point of view so that we prefer to rely on our guess in order not to underate the ability of the data to constrain the variation of $g_{i}(r_0)$.

Should the acceleration scales $g_i(r_0)$ be universal quantities as argued by G09, one should find no correlation with any stellar or DM property. In order to check their hypothesis, we therefore fit the following log\,-\,linear relations

\begin{displaymath}
\log{g_i(r_0)} = a \log{\left ( \frac{M_{vir}}{10^{11} \ {\rm M_{\odot}}} \right )} + b \ ,
\end{displaymath}

\begin{displaymath}
\log{g_i(r_0)} = a M_B + b \ ,
\end{displaymath}
with $i = \{DM, bar\}$ using the Bayesian method\footnote{As a consistency check, we also use the direct fit method, i.e., we minimize the usual $\chi^2$ neglecting the uncertainties on the fitted quantities and the intrinsic scatter. Since the results on the fit parameters $(a, b)$ turn out to be in good agreement, we will only refer in the following to the results from the Bayesian method. We nevertheless show both best fit models in the relevant figures.} in \cite{Dago}. Note that such an approach allows us to take care of the uncertainties on both fitting variables and also to estimate the intrinsic scatter of the correlation in an unbiased way. The results are summarized in Table\,\ref{tab: fitscale} for three different galaxy subsamples.

\begin{figure*}
\centering
\subfigure{\includegraphics[width=7.5cm]{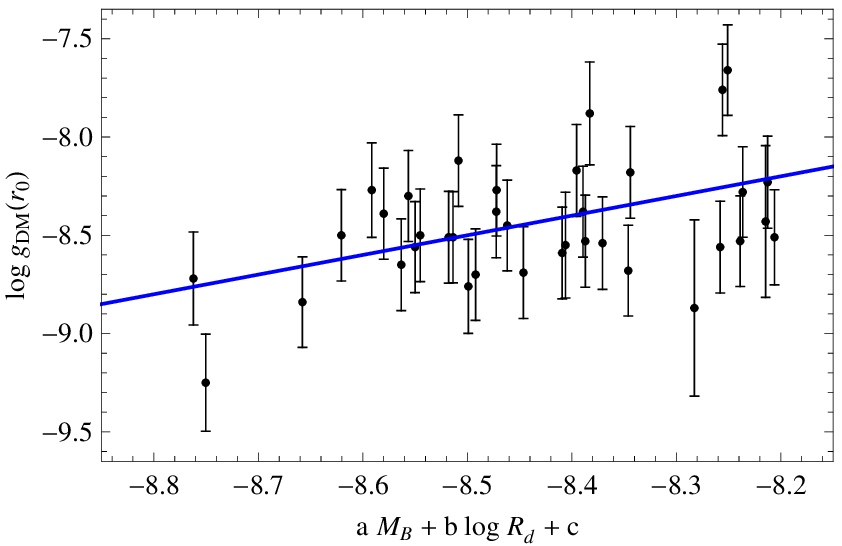}} \goodgap
\subfigure{\includegraphics[width=7.5cm]{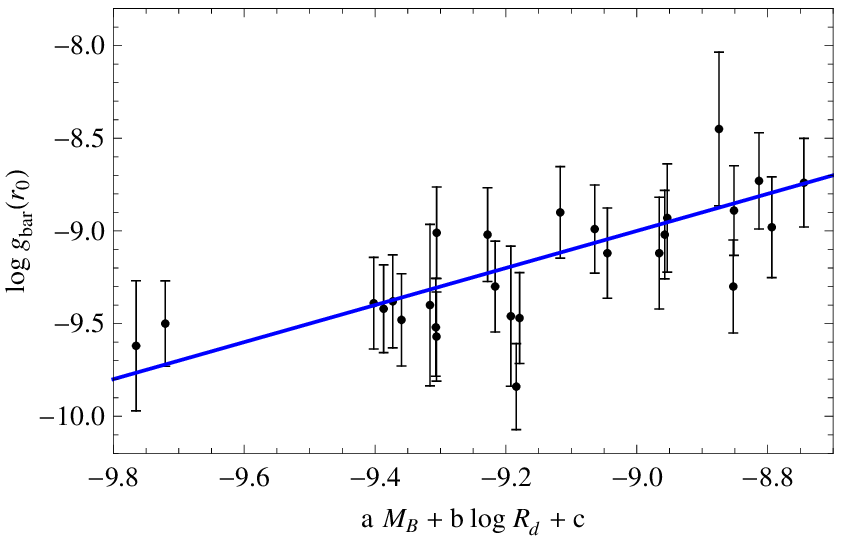}} \goodgap
\caption{A fundamental plane\,-\,like relation for $g_i(r_0)$ as function of $M_B$ and $R_d$.}
\label{fig: gdmbfp}
\end{figure*}

Let us first consider the Good sample made out of all the galaxies well fitted by the Burkert profile and with a scale radius $r_0$ smaller than the radius of the last measured point on the rotation curve. As it is apparent from both the slope values and Fig.\,\ref{fig: gdmfigs}, the hypothesis that the DM acceleration $g_{DM}(r_0)$ is a universal quantity is not supported by the data. Indeed, from a statistical point of view, the absence of any correlation (whether be it with the halo mass $M_{vir}$ or the galaxy luminosity $M_B$) can be safely excluded at the $95\% \ {\rm CL}$. Such a result could seem in fierce disagreement with the G09 one. Actually, this is not the case. Indeed, fitting their smaller galaxy sample, G09 obtained a slope $a \simeq 0.0 \pm 0.2$ for the $g_{DM}(r_0)$ vs $M_B$ so that, although the median value is different, the two results are statistically consistent. Moreover, the intercept of our $g_{DM}(r_0)$ vs $M_B$ correlation translates into $g_{DM}(r_0) = 5.0_{-2.4}^{+4.3} \times 10^{-10} \ {\rm cm/s^2}$ which is well consistent with the claimed constant value found by G09 and quoted before. Given the almost flat slope and the smaller sample used by G09, it is not surprising that their median value for $a$ shifts towards a null one so that we argue that there is no significant difference among the two results of the $g_{DM}(r_0)$ vs $M_B$ relation. What is, nevertheless, different is their interpretation. Indeed, the weak correlation between $g_{DM}(r_0)$ and $M_B$ can be qualitatively considered as a lack of correlation thus motivating the status of a universal quantity given to $g_{DM}(r_0)$ in G09. However, the significant correlation with the halo virial mass can be taken as strong evidence against the universality of $g_{DM}(r_0)$ even if one neglects the very weak but yet statistically meaningful scaling with $M_B$. We therefore conclude that no DM universal acceleration scale may be defined.

While we substantially agree with G09 concerning the scaling of $g_{DM}(r_0)$ with $M_B$, there is on the contrary a rather strong disagreement for the scaling of the baryon acceleration $g_{bar}(r_0)$. In their paper, G09 claim that this quantity is constant over 14 units of $B$ mag, while Fig.\,\ref{fig: gbarfigs} and the slopes in Table\,\ref{tab: fitscale} make us definitely argue against this conclusion. Indeed, a vanishing slope is well outside the $95\%$ CL for both the $g_{bar}(r_0)$\,-\,$M_{vir}$ and $g_{bar}(r_0)$\,-\,$M_B$ relations. Moreover, the $a$ values are not close to zero so that taking a constant $g_{bar}(r_0)$ leads to a severe mismatch to the data.

It is worth wondering why we find a result inconsistent with the G09 ones. To this end, we reduce our sample by only including the 16 galaxies common to both catalogs and fit the scaling relations to this subset. As a result, we find that the slopes of the $g_{DM}(r_0)$\,-\,$M_{vir}$ and $g_{DM}(r_0)$\,-\,$M_B$ relations are now significantly smaller and consistent with null values within the $95\%$ CL so that we recover the G09 result that this quantity can be considered as roughly universal. However, although the scaling laws become shallower, $g_{bar}(r_0)$ still significantly correlates with both the halo virial mass and the galaxy luminosity so that the assumption of constant baryon acceleration can not be retained.

In order to trace the origin of this difference, it is worth first remembering that the G09 sample has been assembled starting from a larger catalog and cutting out objects which do not fulfil some quality selection criteria. On the contrary, the only criteria adopted to include a galaxy in our Good sample relies on the goodness of the fit with the Burkert profile. Should the selection of galaxies be responsible for the different slopes, the fit to the 16 galaxies in common with the two samples would have returned consistent slopes which is not always the case. To further explore this issue, we have plotted in the right panel of Fig.\,\ref{fig: gbarfigs} the G09 $g_{bar}(r_0)$ values superimposed on our data\footnote{This plot also shows us whether the estimated $g_{bar}(r_0)$ values for the galaxies in common to both the G09 and our Good sample agree or not. In the former case, two almost superimposed point will appear in the plot.}. An interesting lesson can be drawn from this figure. Should we exclude the galaxies with $\log{g_{bar}(r_0)} > -8.0$ and add the Milky Way dwarfs with $M_B > -12$, a constant $g_{bar}(r_0)$ would be preferred. All the four systems with $\log{g_{bar}(r_0)} > -8.0$ (namely, NGC\,2841, NGC\,3521, NGC\,4736, NGC\,6946) belong to the THINGS sample so that they have a high quality rotation curve and a not small inclination (as estimated from the HI data). It is, however, nevertheless possible that the presence of warps or non circular motions due to, e.g., bars and/or other sources of instabilities make the rotation curve (notwithstanding the quality of the data) a poor estimator of the underlying gravitational potential thus motivating the exclusion of these systems from the G09 sample. In the range $-20 \le M_B \le -15$, our data are well superimposed on the G09 ones except for few cases where our estimates are larger than those of G09. However, since the values in G09 comes from scaling the result for other halo models to the Burkert one, while our own are directly based on fitting this halo profile to the rotation curve data, we consider our values more reliable. Based on such a comparison, we argue that the most likely reason for the discrepancy on the constancy of $g_{bar}(r_0)$ with $M_B$ is mainly due to our sample including galaxies with larger acceleration values. However, in order to definitely discriminate between the two contrasting results, one should complement our sample with systems probing the luminosity  range $M_B > -12$ to test whether the inferred $g_{bar}(r_0)\,-\,M_B$ scaling can be extrapolated to this regime or a constant value for the baryon acceleration scale should be preferred. It is also worth remembering that $B$ band is a poor estimator of the stellar mass $M_{\star}$ which is actually a far better indicator of the galaxy status. Being $M_{\star}$ hard to estimate in a model independent way, one should better rely whenever possible on {\it Spitzer} $3.6\mu${\it m} magnitudes which are a better tracer of the stellar mass.

As a further test, we have investigated whether the inclusion of systems with poorly determined accelerations can bias the inferred constraints on the scaling relations. Actually, the S/N ratio is typically quite large so that there are no systems prone to large errors. We have nevertheless selected a subsample (referred to as our {\it Gold} sample) requiring the S/N ratio to be larger than 50 for both $g_{i}(r_0)$. Repeating the different fits for this subsample, we find the results in Table\,\ref{tab: fitscale} which are in good agreement with those from the Good sample. We will not discuss further this case in the following, but here we specifically note that the intrinsic scatter decreases for all the correlations investigated.

The scaling relations we have found above are affected by a significant scatter. In order to see whether it is possible to reduce it, we look for a 3D correlation among the acceleration scales and the two stellar parameters directly measurable from the galaxy photometry, namely the total luminosity $M_B$ and the disc scale length radius $R_d$. We therefore use the same Bayesian fitting method to determine the coefficients of the log\,-\,linear relations

\begin{displaymath}
\log{g_i(r_0)} = a M_B + b \log{R_d} + c \ .
\end{displaymath}
Using the Good sample, we find (median and $68\%$ range)\,:

\begin{displaymath}
a = -0.088_{-0.024}^{+0.026} \ \ , \ \ b = -0.072_{-0.042}^{+0.038} \ \ ,
\end{displaymath}

\begin{displaymath}
c = -9.794_{-0.373}^{+0.296} \ \ , \ \ \sigma_{int} = 0.114_{-0.058}^{+0.065} \ \
\end{displaymath}
for $i = DM$ and

\begin{displaymath}
a = -0.271_{-0.040}^{+0.038} \ \ , \ \ b = -0.153_{-0.062}^{+0.066} \ \ ,
\end{displaymath}

\begin{displaymath}
c = -13.46_{-0.46}^{+0.45} \ \ , \ \ \sigma_{int} = 0.246_{-0.067}^{+0.062} \ \
\end{displaymath}
for $i = bar$ (see Fig.\,\ref{fig: gdmbfp}). The addition of a second parameter reduces the intrinsic scatter for both $g_{DM}(r_0)$ and (to a lesser extent) $g_{bar}(r_0)$, but one has also to take into account that a smaller number of galaxies has been used since $R_d$ is not available for all the systems in the Good sample. A definitive conclusion on whether a 2D or 3D scaling law is better suited to forecast the acceleration scales could be possible only increasing the statistics and carefully modeling the uncertainties on the quantities involved.

\section{Conclusions}

Scaling relations among dark matter and stellar quantities can offer valuable insights into the galaxy formation process. Similarly, the existence of universal quantities may be taken as evidence of some hitherto unknown underlying physical mechanism coupling the evolution of the dark matter and baryons in the galaxy. Motivated by the results of D09 and G09 and given the presently confused situation, we have here revisited the issue of the universality of DM and baryon Newtonian acceleration evaluated at the DM scale radius. For this purpose, we have used a large sample of late\,-\,type galaxies spanning roughly nine orders of $B$ magnitude. The use of a MCMC procedure has allowed us to both estimate the model parameters (namely, the stellar $M/L$ ratio and the DM halo virial mass and concentration) and then correctly propagate the errors from both the rotation curve and model parameters to the final estimates of the acceleration scales $g_i(r_0)$ (with $i = DM, bar$ for DM and baryons, respectively).

Using a sample made out of well fitted rotation curve galaxies, we find that $g_{DM}(r_0)$ weakly correlates with the $B$ band absolute magnitude $M_B$ in satisfactorily good agreement with what is expected from the claimed constancy of $\rho_0 r_0$ found by D09. However, we also find a strong correlation with the virial mass $M_{vir}$ which makes us argue against the universality of $g_{DM}(r_0)$. The slopes of the $g_{bar}(r_0)$ vs $M_B$ and $M_{vir}$ relations turn out to be definitely non vanishing so that we can exclude at $95\%$ CL that this quantity is a universal constant. This result is at odds with the G09 one so that it is worth investigating the cause. Compared to G09, there are three main differences. First, we have estimated $g_{bar}(r_0)$ starting from the model parameters obtained by fitting the rotation curve of each galaxy rather than scaling the results in literature to the Burkert profile. While the scaling adopted by G09 typically works satisfactorily well (and, indeed, for the galaxies in the G09 sample we do not find a significant discrepancy), it nevertheless shifts some galaxies in the vertical direction thus possibly affecting the slope estimate. Moreover, our sample covers a smaller $M_B$ range lacking low luminosity systems. Indeed, we find that the points missing in our sample, but present in the G09 one severely departs from the extrapolation of our $g_{bar}(r_0)$\,-\,$M_B$ relation. Finally, we note that, while we only ask that a galaxy is well fitted by the Burkert model to include it in the sample, G09 applied some quality cuts before deriving their results on the constancy of $g_{bar}(r_0)$. Remarkably, if we apply their same cut, we find smaller slopes in marginal agreement with the G09 result. Such a circumstance makes us point at selection effects as a possible drawback of this kind of analysis. On the one hand, simulated samples may be used to fully mimic what is done with actual data and then investigate whether quality cuts may bias the inferred slope. On the other hand, one should increase the sample in order to directly check how the $g_{i}(r_0)$\,-\,$M_B$ and $g_{i}(r_0)$\,-\,$M_{vir}$ relations change depending on the selection cuts applied.

As a final remark, we want to stress that it is worth reconsidering the choice of the reference radius in this kind of analysis. As Fig.\,\ref{fig: dmaccratio} shows, the estimated $g_{i}(R)$ are model independent only if the reference radius $R$ is carefully chosen. From this point of view, the scale radius $r_0$ is not an optimal choice since it probes the acceleration profile in the region where it is most model dependent if the halo concentration is too large. A preliminary study looking for a truly model independent quantity is therefore welcome before trying to understand whether the data argue in favour or against the existence of universal acceleration scale.

\section*{Acknowledgements}

We warmly thank the referee for his/her comments which have helped to significantly improve the manuscript. VFC is funded by the Italian Space Agency (ASI) through contract Euclid\,-\,IC (I/031/10/0).

\appendix

\section{Tables}

Below, we report relevant tables for readers interested in checking the analysis presented in this paper. Table\,\ref{tab: fitres} gives the best fit, median and $68\%$ confidence ranges on the fitted parameters $(\Upsilon_{\star}/\Upsilon_{fid}, c_{vir}, V_{vir})$ for all the galaxies in the sample. Note that we do not directly fit the stellar $M/L$ ratio, but rather its value scaled with respect to the fiducial one, $\Upsilon_{fid}$, adopted in the paper where the data are taken from. We also give constraints on the logarithm of the virial mass and the Burkert scale radius $r_0$.

Table\,\ref{tab: gdmres} gives the data of interest for the fits involving the acceleration scales. We divide the table in two sections. First, we give the galaxies in the Good sample, then the ones discarded because of unreliable fit. We typeset in italic those systems belonging to the Gold subset and in italics those common to the G09 sample. We also report two different quality flags. The first one refers to whether the rotation curve fit is reliable or not, while the second one checks whether the inferred $r_0$ is smaller or larger than $R_{max}$, this latter being the last measured point of the rotation curve.

\begin{table*}
\scriptsize
\caption{Galaxy sample and constraints on model parameters. Columns are as follows\,: 1. galaxy id; 2. reference for the rotation curve data according to the following scheme\,: dBB02 = de Blok \& Bosma (1992), S05 = Simon et al. (2005),  Sp08 = Spano et al. (2008), THINGS = de Blok et al. (2008), Oh = Oh et al. (2011), Sw11 = Swaters et al. (2011); 3. reduced $\tilde{\chi}^2 = \chi^2/d.o.f.$ for the best fit model; 4. best fit model parameters, 5.\,-\,7. median and $68\%$ confidence range for the fitted model parameters $(\Upsilon_{\star}/\Upsilon_{fid}, c_{vir}, V_{vir})$; 8., 9. median and $68\%$ confidence range for the halo virial mass $\log{M_{vir}}$ and scale radius $r_0$. Note that, for galaxies in the S05 sample, we have to set the stellar $M/L$ ratio to its fiducial value since we have only data on the DM circular velocity.}
\begin{center}
\begin{tabular}{ccccccccc}
\hline
Id & Ref & $\tilde{\chi}^2$ & ${\bf p}_{bf}$ & $\Upsilon_{\star}/\Upsilon_{fid}$ & $c_{vir}$ & $V_{vir} \ ({\rm km/s})$ & $\log{(M_{vir}/M_{\odot})}$ & $r_0 \ ({\rm kpc})$ \\
\hline
DDO 53 & Oh & 0.29 & (0.66, 18.7, 27.5) & $0.90_{-0.18}^{+0.27}$ & $16.6_{-2.2}^{+2.9}$ & $37.1_{-13.6}^{+38.6}$ & $10.38_{-0.60}^{+0.93}$ & $4.58_{-2.08}^{+5.58}$ \\
DDO 154 & THINGS & 0.63 & (1.24, 23.7, 30.2) & $1.05_{-0.17}^{+0.13}$ & $23.8_{-0.7}^{+0.7}$ & $30.1_{-0.6}^{+0.7}$ & $10.11_{-0.03}^{+0.03}$ & $2.54_{-0.11}^{+0.13}$ \\
Ho I & Oh & 3.41 & (0.65, 48.7, 13.9) & $0.74_{-0.07}^{+0.15}$ & $49.3_{-2.4}^{+2.7}$ & $13.8_{-0.5}^{+0.4}$ & $9.09_{-0.05}^{+0.04}$ & $0.56_{-0.05}^{+0.05}$ \\
Ho II & Oh & 2.85 & (0.65, 30.6, 17.8) & $0.80_{-0.11}^{+0.17}$ & $27.6_{-4.2}^{+5.4}$ & $17.8_{-0.7}^{+0.9}$ & $9.42_{-0.06}^{+0.06}$ & $1.29_{-0.24}^{+0.29}$ \\
IC 2233 & dBB02 & 0.77 & (1.34, 17.8, 99.9) & $1.12_{-0.27}^{+0.19}$ & $19.0_{-1.3}^{+1.8}$ & $84.7_{-13.5}^{+26.0}$ & $11.45_{-0.22}^{+0.35}$ & $6.87_{-0.85}^{+0.72}$ \\
IC 2574 & THINGS & 0.17 & (0.76, 13.1, 66.0) & $0.91_{-0.11}^{+0.17}$ & $12.4_{-0.6}^{+0.5}$ & $69.1_{-4.6}^{+6.6}$ & $11.19_{-0.09}^{+0.12}$ & $10.92_{-1.04}^{+1.15}$ \\
M81 dwB & Oh & 0.08 & (1.34, 68.8, 15.9) & $1.00_{-0.25}^{+0.25}$ & $69.3_{-5.5}^{+6.4}$ & $16.2_{-1.3}^{+1.6}$ & $9.30_{-0.11}^{+0.12}$ & $0.47_{-0.07}^{+0.09}$ \\
NGC 925 & THINGS & 1.31 & (0.75, 13.8, 111.3) & $0.76_{-0.01}^{+0.01}$ & $13.6_{-0.4}^{+0.5}$ & $115.2_{-9.4}^{+14.2}$ & $11.85_{-0.11}^{+0.15}$ & $13.43_{-0.31}^{+0.63}$ \\
NGC 2366 & THINGS & 0.14 & (0.77, 28.2, 31.5) & $0.94_{-0.13}^{+0.19}$ & $27.4_{-1.7}^{+1.7}$ & $32.0_{-1.7}^{+2.2}$ & $10.18_{-0.07}^{+0.09}$ & $2.34_{-0.26}^{+0.30}$ \\
NGC 2403 & THINGS & 0.93 & (1.25, 28.6, 80.1) & $1.25_{-0.01}^{+0.01}$ & $28.5_{-0.2}^{+0.2}$ & $80.5_{-0.4}^{+0.2}$ & $11.39_{-0.01}^{+0.01}$ & $5.66_{-0.06}^{+0.08}$ \\
NGC 2841 & THINGS & 1.61 & (0.75, 48.5, 143.1) & $0.76_{-0.01}^{+0.03}$ & $48.5_{-1.3}^{+0.4}$ & $142.7_{-0.7}^{+1.1}$ & $12.13_{-0.01}^{+0.01}$ & $5.91_{-0.08}^{+0.17}$ \\
NGC 2976 & THINGS & 0.51 & (0.75, 25.5, 673.9) & $0.77_{-0.01}^{+0.02}$ & $26.0_{-1.1}^{+1.1}$ & $149.4_{-51.4}^{+156.4}$ & $12.19_{-0.55}^{+0.93}$ & $11.63_{-4.27}^{+12.75}$ \\
NGC 3031 & THINGS & 3.58 & (0.78, 39.4, 79.4) & $0.79_{-0.03}^{+0.03}$ & $38.1_{-2.8}^{+2.4}$ & $79.5_{-1.1}^{+1.9}$ & $11.37_{-0.02}^{+0.03}$ & $4.22_{-0.32}^{+0.34}$ \\
NGC 3198 & THINGS & 2.09 & (1.00, 17.9, 90.9) & $0.97_{-0.15}^{+0.11}$ & $18.4_{-2.1}^{+3.1}$ & $90.6_{-2.5}^{+3.0}$ & $11.54_{-0.04}^{+0.04}$ & $9.81_{-1.55}^{+1.71}$ \\
NGC 3274 & dBB02 & 0.42 & (1.32, 62.1, 37.7) & $0.91_{-0.18}^{+0.27}$ & $67.3_{-4.5}^{+4.8}$ & $37.3_{-0.9}^{+1.0}$ & $10.38_{-0.03}^{+0.04}$ & $1.11_{-0.09}^{+0.11}$ \\
NGC 3521 & THINGS & 4.38 & (0.65, 38.3, 87.8) & $0.65_{-0.01}^{+0.01}$ & $37.9_{-1.3}^{+1.3}$ & $88.8_{-2.3}^{+1.9}$ & $11.51_{-0.03}^{+0.03}$ & $4.71_{-0.24}^{+0.23}$ \\
NGC 3621 & THINGS & 0.75 & (1.08, 17.4, 98.9) & $1.05_{-0.02}^{+0.03}$ & $18.1_{-0.9}^{+0.3}$ & $97.5_{-1.4}^{+2.9}$ & $11.64_{-0.02}^{+0.04}$ & $10.84_{-0.26}^{+0.90}$ \\
NGC 4395 & dBB02 & 0.62 & (1.35, 44.4, 36.6) & $1.00_{-0.21}^{+0.26}$ & $45.1_{-2.2}^{+1.5}$ & $37.6_{-1.4}^{+1.6}$ & $10.39_{-0.04}^{+0.06}$ & $1.67_{-0.13}^{+0.15}$ \\
NGC 4455 & dBB02 & 0.22 & (0.92, 24.6, 44.2) & $0.96_{-0.20}^{+0.21}$ & $24.2_{-2.5}^{+2.3}$ & $45.1_{-4.6}^{+6.2}$ & $10.63_{-0.14}^{+0.17}$ & $3.73_{-0.60}^{+0.95}$ \\
NGC 4605 & S05 & 1.56 & (1.00, 62.8, 41.6) & --------- & $62.9_{-1.4}^{+1.5}$ & $41.5_{-1.5}^{+1.6}$ & $10.52_{-0.04}^{+0.05}$ & $1.32_{-0.07}^{+0.08}$ \\
NGC 4736 & THINGS & 1.42 & (0.75, 69.7, 36.3) & $0.89_{-0.09}^{+0.08}$ & $43.3_{-16.7}^{+18.5}$ & $33.6_{-2.5}^{+2.6}$ & $10.25_{-0.10}^{+0.10}$ & $1.50_{-0.37}^{+0.75}$ \\
NGC 5023 & dBB02 & 0.22 & (0.65, 42.9, 46.1) & $0.82_{-0.12}^{+0.20}$ & $40.3_{-2.6}^{+2.6}$ & $47.3_{-2.1}^{+2.1}$ & $10.69_{-0.05}^{+0.06}$ & $2.36_{-0.23}^{+0.26}$ \\
NGC 5055 & THINGS & 3.63 & (0.81, 11.5, 129.8) & $0.81_{-0.01}^{+0.01}$ & $11.3_{-0.4}^{+0.6}$ & $132.1_{-5.6}^{+2.9}$ & $12.03_{-0.06}^{+0.03}$ & $23.66_{-2.17}^{+1.30}$ \\
NGC 5949 & S05 & 0.25 & (1.00, 43.9, 43.5) & --------- & $43.9_{-2.5}^{+3.0}$ & $43.6_{-4.1}^{+4.8}$ & $10.59_{-0.13}^{+0.13}$ & $1.99_{-0.29}^{+0.35}$ \\
NGC 5963 & S05 & 1.78 & (1.00, 71.8, 50.6) & --------- & $71.5_{-3.4}^{+4.0}$ & $50.6_{-2.2}^{+2.6}$ & $10.78_{-0.06}^{+0.07}$ & $1.42_{-0.13}^{+0.15}$ \\
NGC 6689 & S05 & 0.44 & (1.00, 36.6, 64.1) & --------- & $36.7_{-2.1}^{+2.5}$ & $63.1_{-5.7}^{+9.0}$ & $11.07_{-0.08}^{+0.17}$ & $3.46_{-0.49}^{+0.70}$ \\
NGC 6946 & THINGS & 1.07 & (0.65, 30.1, 98.8) & $0.69_{-0.03}^{+0.05}$ & $28.0_{-2.4}^{+1.4}$ & $100.8_{-1.9}^{+4.1}$ & $11.68_{-0.03}^{+0.05}$ & $7.22_{-0.42}^{+1.00}$ \\
NGC 7331 & THINGS & 0.30 & (0.70, 19.1, 128.0) & $0.71_{-0.03}^{+0.02}$ & $18.4_{-1.6}^{+2.0}$ & $131.0_{-7.5}^{+9.7}$ & $12.02_{-0.08}^{+0.09}$ & $14.22_{-2.06}^{+2.50}$ \\
NGC 7793 & THINGS & 2.54 & (1.15, 35.0, 61.9) & $1.13_{-0.04}^{+0.05}$ & $35.0_{-1.1}^{+1.1}$ & $62.2_{-1.5}^{+1.0}$ & $11.05_{-0.03}^{+0.02}$ & $14.22_{-2.06}^{+2.50}$ \\
UGC 731 & Sw11 & 0.88 & (1.33, 46.8, 36.4) & $0.91_{-0.19}^{+0.32}$ & $44.5_{-2.3}^{+2.0}$ & $36.6_{-1.0}^{+1.1}$ & $10.36_{-0.04}^{+0.04}$ & $1.65_{-0.11}^{+0.14}$ \\
UGC 1230 & dBB02 & 0.28 & (0.67, 34.8, 62.8) & $0.85_{-0.15}^{+0.34}$ & $34.0_{-3.6}^{+4.2}$ & $62.3_{-3.8}^{+7.6}$ & $11.05_{-0.08}^{+0.08}$ & $3.67_{-0.55}^{+0.60}$ \\
UGC 1281 & dBB02 & 0.29 & (0.65, 22.4, 45.7) & $0.80_{-0.11}^{+0.23}$ & $21.2_{-2.0}^{+1.8}$ & $48.1_{-6.0}^{+10.2}$ & $10.72_{-0.18}^{+0.25}$ & $4.55_{-0.86}^{+1.50}$ \\
UGC 3137 & dBB02 & 3.46 & (0.65, 16.2, 80.0) & $0.74_{-0.07}^{+0.11}$ & $16.3_{-0.7}^{+0.7}$ & $79.5_{-2.1}^{+2.6}$ & $11.37_{-0.04}^{+0.03}$ & $9.78_{-0.59}^{+0.74}$ \\
UGC 3371 & dBB02 & 0.01 & (0.67, 20.8, 60.5) & $0.72_{-0.18}^{+0.30}$ & $19.7_{-2.4}^{+2.3}$ & $63.0_{-6.8}^{+10.1}$ & $11.07_{-0.15}^{+0.19}$ & $6.40_{-1.25}^{+1.92}$ \\
UGC 4173 & dBB02 & 0.05 & (0.66, 12.1, 36.8) & $0.94_{-0.20}^{+0.26}$ & $10.4_{-2.2}^{+2.9}$ & $38.8_{-7.6}^{+17.7}$ & $10.44_{-0.29}^{+0.49}$ & $6.82_{-6.14}^{+3.49}$ \\
UGC 4256 & Sp08 & 0.66 & (0.66, 36.2, 59.7) & $1.20_{-0.80}^{+3.83}$ & $33.5_{-15.6}^{+5.7}$ & $56.6_{-26.5}^{+5.0}$ & $10.92_{-0.82}^{+0.11}$ & $3.36_{-0.60}^{+0.92}$ \\
UGC 4325 & dBB02 & 0.01 & (0.66, 38.9, 90.3) & $0.92_{-0.20}^{+0.26}$ & $36.0_{-2.8}^{+3.4}$ & $110.6_{-28.1}^{+79.7}$ & $11.80_{-0.38}^{+0.71}$ & $6.05_{-1.78}^{+5.32}$ \\
UGC 4499 & Sw11 & 0.23 & (0.66, 29.1, 41.7) & $0.87_{-0.16}^{+0.27}$ & $28.0_{-2.4}^{+2.0}$ & $41.7_{-1.9}^{+2.3}$ & $10.53_{-0.06}^{+0.07}$ & $3.00_{-0.32}^{+0.43}$ \\
UGC 5414 & Sw11 & 0.27 & (0.66, 27.9, 36.7) & $0.93_{-0.20}^{+0.28}$ & $26.4_{-2.7}^{+2.4}$ & $37.0_{-3.9}^{+5.6}$ & $10.38_{-0.14}^{+0.18}$ & $2.80_{-0.47}^{+0.78}$ \\
UGC 5721 & Sp08 & 8.87 & (0.01, 71.2, 36.1) & $0.36_{-0.25}^{+0.48}$ & $70.9_{-1.0}^{+1.3}$ & $36.1_{-0.5}^{+0.5}$ & $10.34_{-0.02}^{+0.03}$ & $1.02_{-0.03}^{+0.03}$ \\
UGC 6446 & Sw11 & 0.54 & (1.32, 35.4, 40.1) & $0.93_{-0.19}^{+0.25}$ & $38.5_{-2.7}^{+2.4}$ & $40.1_{-1.1}^{+1.1}$ & $10.48_{-0.04}^{+0.04}$ & $2.10_{-0.17}^{+0.19}$ \\
UGC 7323 & Sw11 & 0.22 & (1.35, 22.8, 57.2) & $0.96_{-0.22}^{+0.29}$ & $25.2_{-2.9}^{+2.7}$ & $56.4_{-6.6}^{+10.6}$ & $10.92_{-0.16}^{+0.23}$ & $4.46_{-0.86}^{+1.46}$ \\
UGC 7399 & Sw11 & 3.99 & (1.33, 52.0, 53.1) & $1.21_{-0.24}^{+0.11}$ & $52.4_{-1.4}^{+2.1}$ & $53.1_{-0.7}^{+0.7}$ & $10.84_{-0.02}^{+0.02}$ & $2.03_{-0.10}^{+0.09}$ \\
UGC 7524 & Sw11 & 0.30 & (1.35, 32.2, 43.6) & $1.02_{-0.22}^{+0.24}$ & $32.6_{-1.3}^{+1.2}$ & $44.3_{-1.3}^{+1.3}$ & $10.61_{-0.04}^{+0.04}$ & $2.73_{-0.15}^{+0.18}$ \\
UGC 7559 & Sw11 & 0.06 & (0.74, 39.6, 17.9) & $0.91_{-0.19}^{+0.28}$ & $38.5_{-4.8}^{+5.4}$ & $18.2_{-2.1}^{+2.9}$ & $9.45_{-0.16}^{+0.19}$ & $0.95_{-0.21}^{+0.30}$ \\
UGC 7577 & Sw11 & 0.21 & (0.65, 3.7, 24.5) & $0.75_{-0.07}^{+0.12}$ & $4.3_{-3.0}^{+5.1}$ & $7.8_{-5.2}^{+13.8}$ & $8.35_{-1.44}^{+1.33}$ & $5.15_{-4.3}^{+21.2}$ \\
UGC 7603 & Sw11 & 0.12 & (0.93, 34.5, 35.3) & $0.93_{-0.21}^{+0.25}$ & $34.6_{-2.7}^{+2.4}$ & $35.4_{-1.6}^{+1.7}$ & $10.32_{-0.06}^{+0.06}$ & $2.06_{-0.21}^{+0.27}$ \\
UGC 8490 & Sw11 & 0.69 & (1.35, 51.3, 39.8) & $1.11_{-0.27}^{+0.17}$ & $54.8_{-3.3}^{+3.0}$ & $39.5_{-0.7}^{+0.7}$ & $10.46_{-0.02}^{+0.02}$ & $1.44_{-0.10}^{+0.12}$ \\
UGC 9179 & Sp08 & 0.65 & (1.04, 48.7, 41.4) & $0.96_{-0.52}^{+0.44}$ & $49.3_{-2.8}^{+3.2}$ & $41.2_{-0.7}^{+0.8}$ & $10.51_{-0.02}^{+0.02}$ & $1.68_{-0.13}^{+0.12}$ \\
UGC 9211 & Sw11 & 0.05 & (1.02, 34.0, 36.7) & $0.88_{-0.17}^{+0.29}$ & $34.1_{-2.9}^{+3.3}$ & $36.6_{-1.6}^{+2.2}$ & $10.36_{-0.06}^{+0.08}$ & $2.14_{-0.25}^{+0.34}$ \\
UGC 9465 & Sp08 & 1.30 & (1.08, 17.5, 142.9) & $1.02_{-0.27}^{+0.17}$ & $18.0_{-1.8}^{+3.0}$ & $132.9_{-29.4}^{+46.2}$ & $12.04_{-0.32}^{+0.39}$ & $10.34_{-1.98}^{+1.90}$ \\
UGC 10075 & Sp08 & 2.44 & (0.58, 60.8, 79.0) & $0.64_{-0.28}^{+0.28}$ & $60.4_{-2.4}^{+2.1}$ & $78.7_{-1.2}^{+1.3}$ & $11.36_{-0.02}^{+0.02}$ & $2.62_{-0.06}^{+0.07}$ \\
UGC 10310 & dBB02 & 0.15 & (0.65, 21.5, 47.6) & $0.87_{-0.16}^{+0.28}$ & $19.0_{-3.1}^{+3.7}$ & $50.5_{-8.2}^{+14.3}$ & $10.78_{-0.23}^{+0.32}$ & $5.31_{-1.50}^{+2.60}$ \\
UGC 11557 & Sp08 & 1.65 & (0.04, 25.6, 57.8) & $0.72_{-0.50}^{+0.67}$ & $23.8_{-2.2}^{+2.0}$ & $59.3_{-6.1}^{+8.7}$ & $10.99_{-0.14}^{+0.18}$ & $5.01_{-0.81}^{+1.20}$ \\
UGC 11707 & Sw11 & 0.73 & (1.22, 25.7, 55.3) & $0.86_{-0.17}^{+0.32}$ & $26.0_{-1.5}^{+2.0}$ & $55.7_{-1.9}^{+2.0}$ & $10.91_{-0.04}^{+0.05}$ & $4.30_{-0.43}^{+0.42}$ \\
UGC 12060 & Sw11 & 0.03 & (0.72, 52.0, 37.2) & $0.92_{-0.19}^{+0.26}$ & $51.0_{-4.9}^{+5.1}$ & $36.9_{-1.5}^{+1.5}$ & $10.37_{-0.05}^{+0.05}$ & $1.23_{-0.09}^{+0.05}$ \\
UGC 12632 & Sw11 & 0.29 & (0.65, 38.2, 39.9) & $0.82_{-0.12}^{+0.24}$ & $37.7_{-2.0}^{+2.0}$ & $39.8_{-1.0}^{+1.3}$ & $10.47_{-0.03}^{+0.04}$ & $2.12_{-0.15}^{+0.19}$ \\
UGC 12732 & Sw11 & 1.40 & (1.32, 25.7, 52.8) & $1.02_{-0.24}^{+0.23}$ & $27.0_{-1.6}^{+1.4}$ & $52.4_{-1.4}^{+1.5}$ & $10.83_{-0.03}^{+0.04}$ & $3.89_{-0.28}^{+0.34}$ \\
\hline
\end{tabular}
\end{center}
\label{tab: fitres}
\end{table*}

\begin{table*}
\scriptsize
\caption{Quantities for the acceleration scales fits. Columns are as follows\,: 1. galaxy id; 2. absolute B band magnitude; 3. disc scalelength radius ; 4. RC quality flag (${\cal{Q}}_{RC} = 0$ for bad fit, ${\cal{Q}}_{RC} = 1$ for good fit); 5. $R_0$ quality flag (${\cal{Q}}_0 = 0$ for $r_0 > R_{max}$, ${\cal{Q}}_0 = 1$ for $r_0 \le R_{max}$); 6. - 9. median and $68\%$ confidence range for $\log{g_{i}(r_0)}$ with $i = HI, d, bar, DM$ for gas, disc, baryons and dark matter and the accelerations in ${\rm cm/s^2}$. First part of the table refers to the galaxies in the Good sample, while the second one contains data for rejected systems. We typeset in italics systems belonging to the Gold sample and add a $\star$ symbol to identify those in common with G09.}
\begin{center}
\begin{tabular}{ccccccccc}
\hline
Id & $M_B$ & $R_d \ ({\rm kpc})$ & ${\cal{Q}}_{RC}$ & ${\cal{Q}}_{0}$ & $\log{g_{HI}(r_0)}$ & $\log{g_{d}(r_0)}$ & $\log{g_{bar}(r_0)}$ & $\log{g_{DM}(r_0)}$ \\
\hline
{\it DDO 154}$^{\star}$ & -14.23 & 0.72 & 1 & 1 & $-9.59_{-0.01}^{+0.01}$ & $-10.26_{-0.07}^{+0.05}$ & $-9.50_{-0.01}^{+0.01}$ & $-8.84_{-0.01}^{+0.01}$ \\
{\it Ho I} & -14.80 & ---- & 1 & 1 & $-9.60_{-0.04}^{+0.03}$ & $-10.93_{-0.24}^{+0.15}$ & $-9.58_{-0.03}^{+0.02}$ & $-8.66_{-0.03}^{+0.03}$ \\
Ho II & -16.90 & ---- & 1 & 1 & $-10.28_{-0.30}^{+0.06}$ & $-9.47_{-0.06}^{+0.09}$ & $-9.41_{-0.05}^{+0.08}$ & $-8.96_{-0.11}^{+0.11}$ \\
IC 2233 & -16.61 & 2.30 & 1 & 1 & $-9.32_{-0.58}^{+0.20}$ & $-9.35_{-0.52}^{+0.21}$ & $-9.02_{-0.56}^{+0.18}$ & $-8.57_{-0.02}^{+0.03}$ \\
{\it NGC 2366}$^{\star}$ & -17.17 & ---- & 1 & 1 & $-9.29_{-0.08}^{+0.03}$ & $-9.58_{-0.07}^{+0.08}$ & $-9.11_{-0.05}^{+0.03}$ & $-8.71_{-0.03}^{+0.02}$ \\
{\it NGC 2403}$^{\star}$ & -19.43 & 1.81 & 1 & 1 & $-9.46_{-0.01}^{+0.01}$ & $-8.54_{-0.01}^{+0.01}$ & $-8.49_{-0.01}^{+0.01}$ & $-8.28_{-0.01}^{+0.01}$ \\
{\it NGC 2841} & -21.21 & 4.20 & 1 & 1 & $-10.08_{-0.01}^{+0.01}$ & $-7.80_{-0.01}^{+0.01}$ & $-7.56_{-0.01}^{+0.0}$ & $-7.66_{-0.01}^{+0.01}$ \\
{\it NGC 3198}$^{\star}$ & -20.75 & 3.06 & 1 & 1 & $-9.71_{-0.17}^{+0.06}$ & $-8.39_{-0.05}^{+0.04}$ & $-8.33_{-0.05}^{+0.04}$ & $-8.54_{-0.06}^{+0.09}$ \\
{\it NGC 3521} & -19.88 & ---- & 1 & 1 & $-10.03_{-0.54}^{+0.23}$ & $-7.58_{-0.02}^{+0.01}$ & $-7.58_{-0.02}^{+0.09}$ & $-8.04_{-0.02}^{+0.02}$ \\
{\it NGC 3621}$^{\star}$ & -20.05 & 2.61 & 1 & 1 & $-9.57_{-0.01}^{+0.01}$ & $-8.59_{-0.06}^{+0.02}$ & $-8.55_{-0.05}^{+0.02}$ & $-8.52_{-0.02}^{+0.01}$ \\
{\it NGC 4605} & -17.19 & ---- & 1 & 1 & --------- & --------- & --------- & $-8.00_{-0.01}^{+0.01}$ \\
NGC 4736 & -19.80 & 1.99 & 1 & 1 & $-9.33_{-0.15}^{+0.13}$ & $-7.53_{-0.07}^{+0.04}$ & $-7.17_{-0.23}^{+0.14}$ & $-8.35_{-0.35}^{+0.27}$ \\
{\it NGC 5023} & -16.20 & 0.80 & 1 & 1 & $-9.81_{-0.03}^{+0.09}$ & $-9.89_{-0.08}^{+0.16}$ & $-9.54_{-0.05}^{+0.12}$ & $-8.27_{-0.04}^{+0.04}$ \\
{\it NGC 5055}$^{\star}$ & -21.12 & ---- & 1 & 1 & $-9.72_{-0.11}^{+0.14}$ & $-8.54_{-0.01}^{+0.08}$ & $-8.51_{-0.05}^{+0.09}$ & $-8.70_{-0.02}^{+0.02}$ \\
{\it NGC 5949} & -17.81 & ---- & 1 & 1 & --------- & --------- & --------- & $-8.24_{-0.02}^{+0.02}$ \\
{\it NGC 6689} & -17.43 & ---- & 1 & 1 & --------- & --------- & --------- & $-8.20_{-0.03}^{+0.02}$ \\
{\it NGC 6946} & -20.61 & 2.97 & 1 & 1 & $-9.39_{-0.06}^{+0.05}$ & $-8.02_{-0.04}^{+0.02}$ & $-7.96_{-0.05}^{+0.02}$ & $-8.20_{-0.06}^{+0.03}$ \\
{\it NGC 7331} & -21.67 & ---- & 1 & 1 & $-9.49_{-0.07}^{+0.13}$ & $-8.12_{-0.10}^{+0.11}$ & $-8.04_{-0.09}^{+0.11}$ & $-8.38_{-0.04}^{+0.04}$ \\
{\it UGC 731}  & -16.02 & 1.65 & 1 & 1 & $-10.21_{0.06}^{+0.12}$ & $-9.50_{-0.10}^{+0.12}$ & $-9.42_{-0.08}^{+0.12}$ & $-8.31_{-0.03}^{+0.03}$ \\
{\it UGC 1230} & -17.96 & 4.50 & 1 & 1 & $-9.86_{-0.05}^{+0.04}$ & $-9.92_{-0.10}^{+0.13}$ & $-9.59_{-0.06}^{+0.08}$ & $-8.27_{-0.07}^{+0.07}$ \\
{\it UGC 1281} & -16.72 & 1.70 & 1 & 1 & $-10.10_{-0.04}^{+0.01}$ & $-10.22_{-0.05}^{+0.10}$ & $-9.86_{-0.02}^{+0.04}$ & $-8.71_{-0.03}^{+0.04}$ \\
{\it UGC 3137} & -18.46 & 2.00 & 1 & 1 & $-8.97_{-0.07}^{+0.05}$ & $-9.10_{-0.07}^{+0.09}$ & $-8.73_{-0.07}^{+0.06}$ & $-8.67_{-0.02}^{+0.02}$ \\
{\it UGC 3371} & -17.12 & 3.10 & 1 & 1 & $-9.87_{-0.05}^{+0.17}$ & $-9.88_{-0.16}^{+0.17}$ & $-9.56_{-0.11}^{+0.15}$ & $-8.65_{-0.04}^{+0.04}$ \\
UGC 4173 & -16.33 & 4.50 & 1 & 1 & $-9.80_{-0.31}^{+0.20}$ & $-9.83_{-0.32}^{+0.23}$ & $-9.51_{-0.32}^{+0.21}$ & $-9.25_{-0.09}^{+0.09}$ \\
UGC 4256$^{\star}$ & -21.65 & 5.15 & 1 & 1 & --------- & $-8.95_{-0.46}^{+0.51}$ & $-8.95_{-0.46}^{+0.51}$ & $-8.30_{-0.67}^{+0.10}$ \\
UGC 4325 & -17.76 & 1.60 & 1 & 1 & $-9.23_{-0.06}^{+0.60}$ & $-9.21_{-0.18}^{+0.57}$ & $-8.92_{-0.11}^{+0.58}$ & $-7.97_{-0.08}^{+0.17}$ \\
{\it UGC 4499}$^{\star}$ & -17.40 & 1.49 & 1 & 1 & $-9.48_{-0.10}^{+0.08}$ & $-9.24_{-0.08}^{+0.10}$ & $-9.05_{-0.05}^{+0.08}$ & $-8.59_{-0.04}^{+0.04}$ \\
{\it UGC 5414} & -17.02 & 1.49 & 1 & 1 & $-9.33_{-0.03}^{+0.01}$ & $-9.29_{-0.09}^{+0.12}$ & $-9.01_{-0.05}^{+0.07}$ & $-8.67_{-0.0}^{+0.03}$ \\
{\it UGC 6446} & -17.06 & 1.87 & 1 & 1 & $-10.04_{-0.21}^{+0.15}$ & $-8.94_{-0.09}^{+0.09}$ & $-8.90_{-0.09}^{+0.09}$ & $-8.37_{-0.05}^{+0.04}$ \\
{\it UGC 7323}$^{\star}$ & -18.20 & 2.20 & 1 & 1 & $-9.43_{-0.30}^{+0.09}$ & $-9.04_{-0.10}^{+0.11}$ & $-8.90_{-0.07}^{+0.08}$ & $-8.52_{-0.05}^{+0.04}$ \\
{\it UGC 7524} & -17.74 & 2.58 & 1 & 1 & $-9.76_{-0.10}^{+0.07}$ & $-9.23_{-0.10}^{+0.09}$ & $-9.12_{-0.08}^{+0.08}$ & $-8.45_{-0.02}^{+0.02}$ \\
{\it UGC 7559} & -13.12 & 0.67 & 1 & 1 & $-9.69_{-0.04}^{+0.01}$ & $-9.92_{-0.10}^{+0.12}$ & $-9.47_{-0.05}^{+0.05}$ & $-8.71_{-0.06}^{+0.05}$ \\
{\it UGC 7603} & -15.81 & 0.90 & 1 & 1	& $-9.64_{-0.21}^{+0.21}$ & $-9.14_{-0.09}^{+0.09}$ & $-9.01_{-0.09}^{+0.09}$ & $-8.51_{-0.04}^{+0.04}$ \\
UGC 8490$^{\star}$ & -16.88 & 0.67 & 1 & 1 & $-9.39_{-0.02}^{+0.01}$ & $-9.15_{-0.54}^{+0.16}$ & $-8.95_{-0.28}^{+0.11}$ & $-8.17_{-0.04}^{+0.04}$ \\
{\it UGC 9179}$^{\star}$ & -17.73 & 1.07 & 1 & 1 & $-9.40_{-0.02}^{+0.02}$ & $-9.03_{-0.31}^{+0.12}$ & $-8.87_{-0.20}^{+0.09}$ & $-8.18_{-0.04}^{+0.04}$ \\
{\it UGC 9211} & -15.87 & 1.32 & 1 & 1 & $-10.31_{-0.30}^{+0.29}$ & $-9.58_{-0.09}^{+0.12}$ & $-9.49_{-0.09}^{+0.10}$ & $-8.50_{-0.05}^{+0.05}$ \\
{\it UGC 9465}$^{\star}$ & -18.25 & 2.29 & 1 & 1 & --------- & $-9.26_{-0.12}^{+0.08}$ & $-9.26_{-0.12}^{+0.08}$ & $-8.39_{-0.02}^{+0.03}$ \\
UGC 10075$^{\star}$ & -19.78 & 2.49 & 1 & 1 & --------- & $-8.60_{-0.26}^{+0.16}$ & $-8.60_{-0.26}^{+0.16}$ & $-7.75_{-0.04}^{+0.03}$ \\
UGC 10310 & -16.81 & 1.90 & 1 & 1 & $-9.37_{-0.48}^{+0.09}$ & $-9.46_{-0.45}^{+0.21}$ & $-9.10_{-0.48}^{+0.12}$  & $-8.75_{-0.07}^{+0.06}$ \\
UGC 11557$^{\star}$ & -19.12 & 1.70 & 1 & 1 & $-9.57_{-0.17}^{+0.26}$ & $-9.59_{-0.49}^{+0.27}$ & $-9.23_{-0.25}^{+0.16}$ & $-8.54_{-0.05}^{+0.03}$ \\
{\it UGC 11707} & -17.50 & 4.30 & 1 & 1 & $-10.20_{-0.05}^{+0.06}$ & $-9.51_{-0.09}^{+0.13}$ & $-9.43_{-0.07}^{+0.11}$ & $-8.51_{-0.03}^{+0.04}$ \\
{\it UGC 12060}$^{\star}$ & -16.60 & 1.76 & 1 & 1 & $-10.77_{-0.57}^{+0.51}$ & $-9.02_{-0.11}^{+0.10}$ & $-9.01_{-0.11}^{+0.10}$ & $-8.13_{-0.03}^{+0.04}$ \\
{\it UGC 12632} & -16.52 & 2.57 & 1 & 1 & $-9.87_{-0.02}^{+0.01}$ & $-9.68_{-0.07}^{+0.11}$ & $-9.47_{-0.03}^{+0.08}$ & $-8.39_{-0.03}^{+0.03}$ \\
{\it UGC 12732} & -16.91 & 2.21 & 1 & 1 & $-10.25_{-0.10}^{+0.06}$ & $-9.32_{-0.11}^{+0.08}$ & $-9.27_{-0.10}^{+0.07}$ & $-8.51_{-0.03}^{+0.03}$ \\
\hline
DDO 53 & -13.40 & ---- & 1 & 0 & $-9.33_{-0.42}^{+0.89}$ & $-10.65_{-0.31}^{+0.44}$ & $-9.31_{-0.38}^{+0.87}$ & $-8.99_{-0.14}^{+0.24}$ \\
IC 2574$^{\star}$ & -18.11 & ---- & 1 & 0 & $-9.38_{-0.11}^{+0.32}$ & $-9.85_{-0.08}^{+0.07}$ & $-9.25_{-0.08}^{+0.25}$ & $-8.92_{-0.01}^{+0.02}$ \\
M81 dwB & -14.20 & ---- & 1 & 0 & $-9.10_{-0.02}^{+0.02}$ & $-9.22_{-0.12}^{+0.09}$ & $-8.86_{-0.05}^{+0.05}$ & $-8.33_{-0.04}^{+0.03}$ \\
NGC 925$^{\star}$ & -20.04 & ---- & 0 & 0 & $-9.20_{-0.03}^{+0.03}$ & $-9.15_{0.34}^{+0.13}$ & $-8.87_{-0.13}^{+0.04}$ & $-8.67_{-0.01}^{+0.01}$ \\
NGC 2976 & -17.78 & ---- & 1 & 0 & $-8.27_{-0.44}^{+0.51}$ & $-5.94_{-0.41}^{+0.49}$ & $-5.94_{-0.41}^{+0.50}$ & $-8.08_{-0.16}^{+0.29}$ \\
NGC 3031 & -20.73 & ---- & 0 & 1 & $-9.54_{-0.06}^{+0.07}$ & $-7.58_{-0.02}^{+0.02}$ & $-7.43_{-0.04}^{+0.04}$ & $-8.08_{-0.06}^{+0.04}$ \\
NGC 5963 & -17.35 & ---- & 0 & 1 & --------- & --------- & --------- & $-7.82_{-0.02}^{+0.03}$ \\
NGC 7793 & -18.79 & ---- & 0 & 1 & $-9.34_{-0.05}^{+0.01}$ & $-8.44_{-0.02}^{+0.02}$ & $-8.39_{-0.02}^{+0.02}$ & $-8.25_{-0.02}^{+0.01}$ \\
UGC 5721$^{\star}$ & -16.38 & 0.43 & 0 & 1 & $-9.37_{-0.01}^{+0.01}$ & $-10.07_{-0.52}^{+0.37}$ & $-9.29_{-0.06}^{+0.09}$ &$-7.97_{-0.01}^{+0.01}$ \\
UGC 7399 & -16.01 & 0.79 & 0 & 1 & $-9.41_{-0.01}^{+0.01}$ & $-8.87_{-0.07}^{+0.05}$ & $-8.76_{-0.06}^{+0.03}$ & $-8.03_{-0.01}^{+0.03}$ \\
\hline
\end{tabular}
\end{center}
\label{tab: gdmres}
\end{table*}

\end{document}